\documentclass[12pt,nofootinbib,superscriptaddress,tightenlines,]{revtex4}

\usepackage{amssymb,color,paralist,amsmath,epsfig}
\usepackage{graphicx}
\usepackage{comment}
\usepackage{amsfonts}
\usepackage{relsize}
\usepackage{nicefrac,esint}
\usepackage{float}

\usepackage[colorlinks,citecolor=blue,linktoc=all,linkcolor=cyan]{hyperref}

\def\half{\mbox{\small{$\frac{1}{2}
$}}}

\newcommand{\hp}{{+}}
\newcommand{\hm}{{-}}

\newcommand{\beq}{\begin{equation}}
\newcommand{\eeq}{\end{equation}}

\newcommand{\ber}{\begin{eqnarray}}
\newcommand{\eer}{\end{eqnarray}}

\begin{document}

\title{Forward two-photon exchange in elastic lepton-proton scattering and hyperfine-splitting correction}

\author{Oleksandr Tomalak}
\affiliation{Institut f\"ur Kernphysik and PRISMA Cluster of Excellence, Johannes Gutenberg Universit\"at, Mainz, Germany}


\date{\today}

\begin{abstract}
We relate the forward two-photon exchange (TPE) amplitudes to integrals of the inclusive lepton-proton scattering cross sections. These relations yield an alternative way for the evaluation of the TPE correction to hyperfine-splitting (HFS) in the hydrogen-like atoms with an equivalent to the standard approach (Iddings, Drell and Sullivan) result implying the Burkhardt-Cottingham sum rule. For evaluation of the individual effects (e.g., elastic contribution) our approach yields a distinct result. We compare both methods numerically on examples of the elastic contribution and the full TPE correction to HFS in electronic and muonic hydrogen.
\end{abstract}

\maketitle
\smallskip
\tableofcontents

\section{Introduction}

The two-photon exchange (TPE) corrections could be responsible for the discrepancy in the ratio of the proton electric to magnetic form factors between the traditional Rosenbluth separation \cite{Rosenbluth:1950yq} and the polarization transfer  \cite{Jones:1999rz} methods, cf.~\cite{Guichon:2003qm,Blunden:2003sp} and \cite{Carlson:2007sp,Arrington:2011dn} for reviews. These corrections introduce the largest hadronic uncertainty in the elastic electron-proton scattering experiments and should be precisely accounted for in analysis of the modern data. Dedicated experiments to measure the TPE correction from the ratio of the elastic electron-proton to positron-proton scattering cross sections, where TPE enters with different signs, have recently been carried out at VEPP-3 (Novosibirsk) \cite{Rachek:2014fam}, OLYMPUS (DESY) \cite{Henderson:2016dea}, and CLAS (JLab) \cite{Adikaram:2014ykv}. These measurements show evidence of a significant TPE effect.

The TPE correction also play a prominent role in evaluation of the proton structure (finite-size) contributions to the spectrum of hydrogen (H) and muonic hydrogen ($\mu$H). The uncertainty of TPE is the dominant uncertainty in the precise spectroscopy measurements with the muonic hydrogen. These corrections received a renewed attention in light of the so-called {\it proton radius puzzle}, the discrepancy in the extracted  proton charge radius from the muonic hydrogen Lamb shift \cite{Pohl:2010zza,Antognini:1900ns} and measurements with electrons \cite{Bernauer:2010wm,Bernauer:2013tpr,Mohr:2012tt}, see \cite{Antognini:1900ns,Carlson:2015jba} for recent reviews. One of the recent achievements in this field is the measurement of the 2S hyperfine splitting (HFS) in muonic hydrogen by  the CREMA Collaboration at PSI~\cite{Antognini:1900ns}. Also a high-precision measurement of 1S HFS in muonic hydrogen with an unprecendented ppm accuracy level is planned by this~\cite{Pohl:2016tqq,Dupays:2003zz} and other collaborations~\cite{Ma:2016etb,Adamczak:2016pdb}. Such accurate measurements will strictly constrain the elastic and inelastic proton structure.

\begin{figure}[h]
\begin{center}
\includegraphics[width=.43\textwidth]{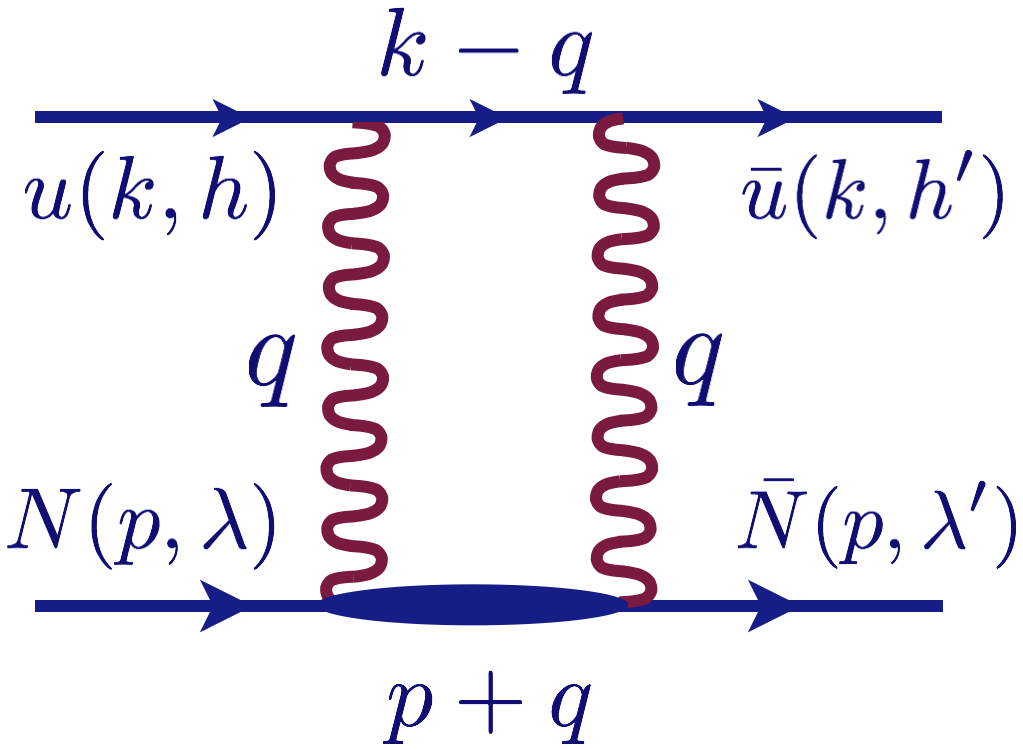}
\end{center}
\caption{Two-photon exchange graph.}
\label{TPE_graph}
\end{figure}

To evaluate the contribution from the diagrams with two exchanged photons to atomic spectrum one notices that the lower part of the TPE graph, see Fig. \ref{TPE_graph}, corresponds with the process of forward doubly virtual Compton scattering (VVCS). The latter in the forward kinematics can be expressed through dispersion relations (DRs) in terms of the proton structure functions (SFs)  measured in elastic and inelastic electron-proton ($ep$) scattering.  The forward TPE is thus evaluated, performing the Wick rotation, as an integral over the photon energy, $ \nu_\gamma $, and virtuality, $Q^2 > 0$. For the Lamb shift correction evaluation \cite{Pachucki:1996zza,Faustov:1999ga,Pineda:2002as,Pineda:2004mx,Nevado:2007dd,Carlson:2011zd} the subtraction function is needed in addition. However, the latter can be estimated at low virtualities from the chiral perturbation theory \cite{Birse:2012eb,Alarcon:2013cba,Peset:2014jxa} or non-relativistic quantum electrodynamics (NRQED) \cite{Hill:2012rh} and at high virtualities from the operator product expansion \cite{Hill:2016bjv}. This function, in principle, can also be determined with account of the high-energy SFs data~\cite{Gorchtein:2013yga,Tomalak:2015hva,Caprini:2016wvy}. The leading in the electromagnetic coupling constant TPE effects of the proton structure in HFS can be entirely expressed in terms of SFs exploiting the formalism of the projection operators on the singlet and triplet states \cite{Zemach:1956zz,Iddings:1959zz,Iddings:1965zz,Drell:1966kk,Faustov:1966,Faustov:1970,Bodwin:1987mj,Faustov:2001pn,Carlson:2008ke,Carlson:2011af}. For recent numerical evaluations of the TPE correction to HFS see Refs.~\cite{Faustov:2001pn,Carlson:2008ke,Carlson:2011af}, for the model-independent evaluation within the frameworks of NRQED and ChPT exploiting the hydrogen HFS measurement see Ref. \cite{Peset:2016wjq} and for results in chiral EFT see Ref.~\cite {Hagelstein:2015egb}.

In this work we provide a different way to express the forward TPE contributions in terms of the proton SFs.  We are working on the level of the forward lepton-proton scattering amplitudes and account for the forward double spin-flip amplitude (i.e., the one where the helicities of both lepton and proton are flipped) for the first time. The method we are using to derive these relations is akin to deriving sum rules for Compton or light-by-light scattering \cite{GellMann:1954db,Pascalutsa:2010sj}. For the scattering of two charged particles the derivation changes quite a bit. The crossing relates the amplitude of the particle scattering with the amplitude of the antiparticle scattering \cite{Grein:1977mn,Kroll:1973re}. Nevertheless, for the TPE amplitudes we exploit the crossing in one channel due to the charge independence of the TPE contributions. In this way our dispersion relations (DRs) for 
the elastic lepton-proton scattering do not involve the crossed channels, and hence are different from generic DRs for charged particles, such as the DRs for the nucleon-nucleon scattering \cite{Grein:1977mn,Kroll:1973re,Guichon:1982eb}. Advantageously, the method based on DRs for $ l p $ amplitudes \cite{Gorchtein:2006mq,Borisyuk:2008es,Tomalak_PhD} does not require one to evaluate the poles contributions arising after the Wick rotation for amplitudes above the threshold.

We express the TPE corrections to the Lamb shift and HFS of S energy levels through the forward TPE amplitudes at threshold. With DRs for the lepton-proton amplitudes we are not able to express the correction to the Lamb shift through the experimental information, but the correction to the hyperfine splitting is entirely expressed through the proton spin SFs. The resulting HFS correction agrees with the standard approach of Iddings, Drell and Sullivan {\it et
al.}~\cite{Iddings:1959zz,Iddings:1965zz,Drell:1966kk,Faustov:1966,Faustov:1970,Bodwin:1987mj} only after account for the Burkhardt-Cottingham (BC) sum rule \cite{Burkhardt:1970ti}. However, the contribution of each individual channel to the TPE correction in this work differs from the literature result, which can be also obtained exploiting the DRs for the VVCS amplitudes. Afterwards, we compare the proton state and total TPE correction to HFS in $ \mathrm{H} $ and $ \mu \mathrm{H}$ exploiting the DRs for the lepton-proton amplitudes and for the VVCS amplitudes. We connect the region with small photons virtualities in HFS integrand that we express in terms of moments of the spin SFs to the region with large photons virtualities, where the two methods give the same integrands.

The paper is organized as follows. We write down DRs for the forward elastic $ lp $ scattering TPE amplitudes in terms of the inclusive $lp$ cross sections at leading $ \alpha $ order in Sec. \ref{sum_rules} and verify them in QED in App. \ref{verification}. We relate the inclusive $lp$ cross sections and the forward TPE amplitudes to proton SFs in Sec. \ref{inelastic_lp}. Subsequently, we derive the leading in the electromagnetic coupling constant proton structure corrections to S energy levels coming from TPE and compare the unitarity-based method with the standard approach in Sec. \ref{hfs_correction}. The numerical comparison of the proton and inelastic intermediate states TPE contributions to HFS within two methods is given in Sec. \ref{hfs_correction_numbers}. We give our conclusions in Sec. \ref{conclusions}. We also provide the derivation of the TPE correction to HFS using the forward Compton scattering tensor in App. \ref{box_graph_method}. In App. \ref{scattering_observables} we classify all possible elastic scattering polarization observables in the forward kinematics.

\section{Forward lepton-proton scattering}

The forward elastic $lp$ scattering is described by three \cite{Gribov:1963gx} non-vanishing independent helicity amplitudes $ T_{h^\prime \lambda^\prime h \lambda} $, with $h$ ($h^\prime$) and $\lambda$ ($\lambda^\prime$) the helicities of the inital (final) lepton and proton, see Fig. \ref{TPE_graph} for the notations of kinematics and helicities: 
\ber
 T_{\hp \hp \hp \hp}, \quad T_{\hp \hm\hp \hm} , \quad T_{\hm \hm \hp \hp} , 
\eer
which are functions of the lepton energy $\omega$ in the lab frame. We denote the positive helicity as $ + $ and the negative helicity as $ - $. The contribution with exchange of a fixed number of photons to the following three amplitudes has definite even-odd properties with respect to the crossing $\omega \to -\omega$:
\begin{subequations}
\ber \label{forward_stramp}
f_\pm \left(\omega \right) & = & \half \big(T_{\hp \hp \hp \hp} \pm T_{\hp \hm \hp \hm} \big),  \\
g \left(\omega \right) & = & \half T_{\hm \hm \hp \hp}.
\eer
\end{subequations}
The Lorentz structure of the forward amplitude is then given by
\ber \label{covariant_expression}
T_{h^\prime \lambda^\prime, h \lambda}  \left(\omega \right) & = & \frac{f_+ \left(\omega \right)}{4 m M} \, \bar{u} (k, h^\prime) u (k, h) \,  \bar{N} (p, \lambda^\prime) N (p, \lambda) \nonumber\\
&-& \frac{m f_- \left(\omega \right) + \omega g \left(\omega \right)}{8 M \vec{k}^2} \, \bar{u} (k, h^\prime)\gamma^{\mu \nu} u (k, h) \,  \bar{N} (p, \lambda^\prime) \gamma_{\mu \nu} N (p, \lambda) \nonumber \\
&+& \frac{\omega f_- \left(\omega \right) + m g \left(\omega \right) }{4 M \vec{k}^2 }  \, \bar{u} (k, h^\prime)\gamma_{\mu} \gamma_5 u (k, h)  \, \bar{N} (p, \lambda^\prime) \gamma^{\mu} \gamma_5  N (p, \lambda),
\eer
with $ m $ and $u$  the lepton mass and spinor, $ \vec{k} $ the lepton momentum in the lab frame, $ M $ and $N$ the proton mass and spinor, $\gamma^{\mu \nu} = \frac12[\gamma^\mu,\gamma^\nu]$, where $ \gamma^\mu $ are the Dirac matrices, the spinors are normalized as 
\ber
\bar{u}(k, h^\prime) u (k, h)= 2 m \delta_{h^\prime, h}, ~\bar{N} (p, \lambda^\prime) N (p, \lambda)= 2 M \delta_{\lambda^\prime, \lambda}.
\eer

In order to establish the even-odd properties for the invariant amplitudes under $ \omega \to - \omega $, we first perform the crossing on the lepton line and relate amplitudes of the lepton-proton scattering $ f^{l^- p} (\omega) $ in the physical region ($ \omega > 0$) to amplitudes of the antilepton-proton scattering $ f^{l^+p} (-\omega) $ in the unphysical region ($\omega < 0$):
\ber
f^{ l^+ p}_+\left(\omega  \right) & = & f^{ l^- p}_+\left( - \omega  \right), \\
f^{ l^+ p}_-\left(\omega  \right) & = & - f^{ l^- p}_-\left( - \omega  \right), \\
g^{ l^+ p}\left(\omega  \right) & = & g^{ l^- p}\left( - \omega  \right),
\eer
where $ \omega $ is treated as a complex variable, see Appendix \ref{even_odd} for details of this derivation. The perturbative contributions with odd number of photons connected to the lepton (antilepton) line have different sign in the amplitudes of the lepton-proton and antilepton-proton scattering as compared to the contributions with an even number of photons, which have the same sign. We express the scattering amplitudes in terms of the contributions with even $f^{(2 n) \gamma} \left(\omega  \right)$ and odd $f^{(2 n - 1) \gamma} \left(\omega  \right)$ number of photons connected to both lepton and proton lines, e.g.:
\ber
f^{l^{\pm} p}_- = \sum \limits_{n=1}^{\infty} \left(f^{(2 n) \gamma}_- \pm f^{(2n - 1) \gamma}_- \right),
\eer
and obtain the following crossing relations for the contribution of graphs with $n$ exchanged photons on the real $ \omega $ axis:
\ber
f^{n \gamma}_+\left(\omega  \right) & = & \left( -1 \right)^{n} \left( f^{n \gamma}_+\left(-\omega  \right) \right)^{*}, \label{crossing_relations_TPE_1}  \\
f^{n \gamma}_-\left(\omega  \right) & = & - \left( -1 \right)^{n} \left( f^{n \gamma}_-\left(-\omega  \right)  \right)^{*},\label{crossing_relations_TPE_2}  \\
g^{n \gamma}\left(\omega  \right) & = & \left( -1 \right)^{n} \left( g^{n \gamma}\left(-\omega  \right) \right)^{*}.\label{crossing_relations_TPE_3}
\eer

As usual, the optical theorem establishes the relations between the imaginary parts of the forward amplitudes and the total inclusive cross sections of $lp$ collisions:
\ber
 \Im f_\pm \left(\omega \right) & = &  M |\vec{k}| \left ( \sigma_{\hp \hp}\left(\omega \right) \pm \sigma_{\hp \hm}\left(\omega \right) \right ),  \label{optical_M_1} \\
 \Im g \left(\omega \right) & = &  2 M |\vec{k}|  \left ( \sigma_{\parallel}\left(\omega \right) - \sigma_{\perp}\left(\omega \right) \right ),\label{optical_M_3}
\eer
 where $ \sigma_{h \lambda} $ is the inclusive cross section with the incoming lepton helicity $ h $ and the incoming proton helicity $ \lambda $; $ \sigma_{\perp} $ ($ \sigma_{\parallel} $) is the inclusive cross section with lepton and proton polarized transversely and perpendicular (parallel)  to each other. \footnote{The TPE correction from the inelastic intermediate states was related to the unpolarized photoabsorption cross section in Ref. \cite{Brown:1970te}.}

The elastic (proton) contribution to the inclusive cross section is infrared divergent. This divergence should be subtracted in a proper way in all three amplitudes. We realize this subtraction only for the case of amplitudes at threshold in Section \ref{hfs_correction}.
 
 We express the latter cross sections in terms of the proton SFs up to the order $ \alpha^2 $ in Section \ref{inelastic_lp} and obtain the imaginary parts of the TPE amplitudes at the leading $ \alpha $ order with Eqs. (\ref{optical_M_1}-\ref{optical_M_3}).
 \label{sum_rules}
\begin{figure}[h]
\begin{center}
\includegraphics[width=.45\textwidth]{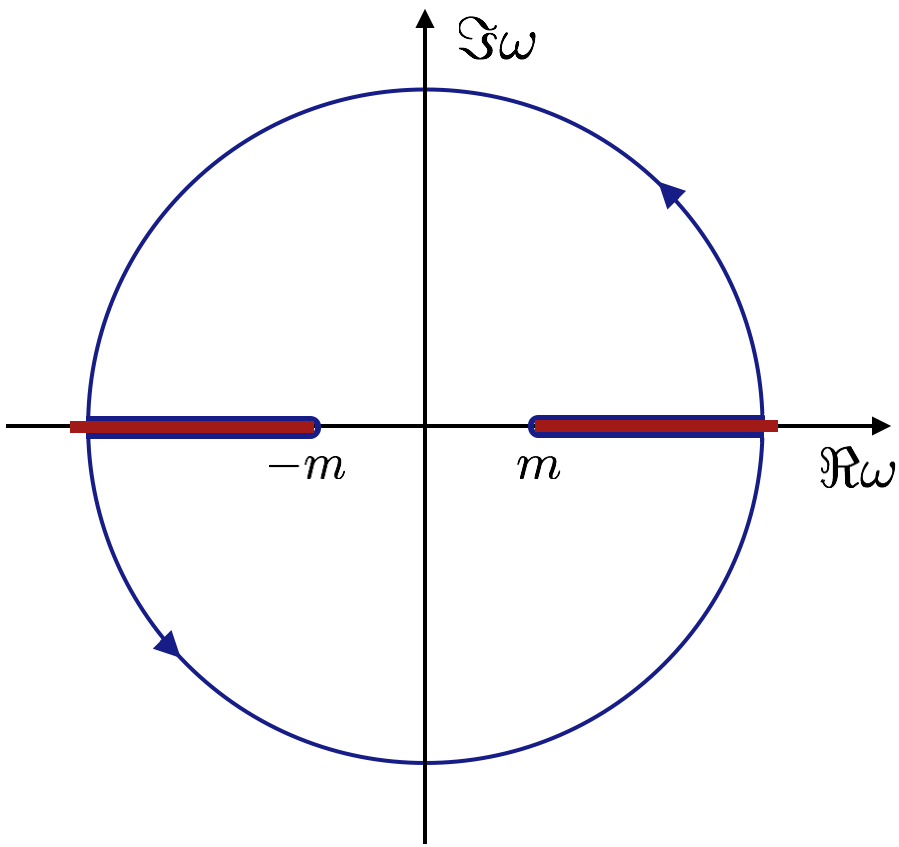}
\end{center}
\caption{Complex plane of the $ \omega $ variable.}
\label{complex_w_plane}
\end{figure}

Assuming analyticity of these amplitudes in the entire complex $\omega$-plane, except for the branch cuts along the real axis extending from threshold to infinity, see Fig. \ref{complex_w_plane}, we can write down the standard DRs:
\ber
 \Re \left\{\begin{array}{l} f_{\pm} \left(\omega \right)\\
g (\omega)\end{array} \right\}  & = & \frac{1}{\pi} \left(\,\fint\limits^{-m}_{-\infty} +\fint\limits^{\infty}_{m} \,\right)\,
  \frac{ \mathrm{d} \omega^\prime }{ \omega^\prime-\omega } \Im \left\{\begin{array}{l} f_{\pm} (\omega^\prime)\\
g (\omega^\prime)\end{array} \right\} ,  \label{Mi_DR} 
\eer
where $\fint$ stands for the principal-value integration. Using properties of the TPE amplitudes under the crossing $ \omega \to -\omega $, see Eqs. (\ref{crossing_relations_TPE_1}-\ref{crossing_relations_TPE_3}), and assuming the sufficient high-energy behavior of the TPE amplitudes in order to neglect the contribution of the contour at infinity in the Cauchy integral formula, we write down DRs valid at the leading $ \alpha $ order and thereby account for both direct and crossed graphs:
\ber 
\Re f^{2 \gamma}_+ \left(\omega \right) & = & \frac{4 M}{\pi} \fint \limits^{ \infty}_{m} \frac{ \omega^\prime  |\vec{k}^\prime| \sigma\left(\omega^\prime \right)}{ \omega^{\prime 2}-\omega^2 }  \mathrm{d} \omega^\prime ,  \label{M1_DR} \\
\Re f^{2 \gamma}_- \left(\omega \right) & = & \frac{2 M \omega }{\pi}  \fint \limits^{ \infty}_{m}  \frac{ |\vec{k}^\prime| \left(\sigma_{\hp \hp}\left(\omega^\prime \right) - \sigma_{\hp \hm}\left(\omega^\prime \right) \right)}{\omega^{\prime 2}-\omega^2}  \mathrm{d} \omega^\prime,  \label{M2_DR} \\ 
\Re g^{2 \gamma} \left(\omega \right) & = & \frac{4 M}{\pi}  \fint \limits^{ \infty}_{m}  \frac{\omega^\prime |\vec{k}^\prime| \left(\sigma_{\parallel}\left(\omega^\prime \right) - \sigma_{\perp}\left(\omega^\prime \right) \right)}{\omega^{\prime 2}-\omega^2}  \mathrm{d} \omega^\prime,  \label{M3_DR}
\eer
with the lepton momentum in the lab frame $| \vec{k}^\prime| = \sqrt{\omega^{\prime 2}-m^2}$. These DRs are written for amplitudes in one channel contrary to the DRs for the forward proton-proton scattering \cite{Grein:1977mn}.

The high-energy behavior of the total unpolarized inclusive cross section does not allow to neglect the contribution of the contour at infinity and to write down Eq. (\ref{M1_DR}). One should perform a subtraction in the DR for the amplitude $ f_+^{2 \gamma} $, e.g., at the point $\omega_s$: \footnote{In the language of effective field theories this subtraction corresponds to the counter term, which was studied in context of the atomic physics and muon-proton scattering in Ref. \cite{Miller:2012ne}.}
\ber \label{fermion_fermion_subtracted}
\Re f^{2 \gamma}_+\left(\omega \right) - \Re f^{2 \gamma}_+\left(\omega_s \right)  =  \frac{4 M \left( \omega^2 - \omega_s^2 \right)}{\pi} \fint \limits^{\infty}_{m} \frac{ \omega^\prime  |\vec{k}^\prime| \sigma \left( \omega^\prime \right)}{\left( \omega^{\prime 2}-\omega^2 \right)\left(\omega^{\prime 2}-\omega_s^2 \right )}  \mathrm{d} \omega^\prime.
\eer
The lepton-proton DRs were checked at the leading QED order, see Appendix \ref{verification} for details. However, the Regge behavior makes the inclusive cross section $ \sigma $ divergent due to the virtual photons with high energy in the lab frame. Consequently the DRs of Eqs. (\ref{M1_DR}) and (\ref{fermion_fermion_subtracted}) cannot be written for the inelastic TPE contribution to the unpolarized amplitude $ f_+ $. 

The DRs of Eqs. (\ref{M2_DR}-\ref{fermion_fermion_subtracted}) have the same form as the DRs for the light-by-light scattering \cite{Pascalutsa:2010sj}.

We decompose the forward scattering amplitudes into a sum of one-photon exchange (OPE) and TPE contributions. The TPE amplitudes, except for $ \Re f^{2 \gamma}_+ $, are obtained with DRs and unitarity relations as described above. The OPE amplitudes are real and given by
\ber
f^{1 \gamma}_- & = & e^2 \mu_P, \\
g^{1 \gamma} & = & 0,
\eer
with the proton magnetic moment $ \mu_P \approx 2.793 $. The vacuum polarization correction is zero in the forward scattering. The lepton vertex correction does not change the amplitude $ g $, modifies the amplitude $ f_- $ by the lepton anomalous magnetic moment $ a_l $: $ \delta f_- =  a_l  e^2$ and contributes to the amplitude $ f_+ $. The proton vertex correction with the proton intermediate state contributes to $ f_- $ and $ f_+ $ amplitudes \cite{Maximon:2000hm}. The contribution of inelastic intermediate states is expected to be small. However, this correction requires an additional theoretical investigation. Therefore, the forward amplitudes are completely expressed in terms of the total inclusive cross sections, see Eqs. (\ref{optical_M_1},~\ref{optical_M_3},~\ref{M2_DR},~\ref{M3_DR}), at $ O (\alpha^2) $ up to the one unknown spin-independent amplitude $ \Re f_+ $.

\section{Relation of the forward two-photon amplitudes to the proton structure}
\label{inelastic_lp}

In this Section, we first express the total inclusive cross sections in terms of the experimentally measured proton SFs. Exploiting these relations, we express the real parts of the forward TPE amplitudes as integrals over the photon energy $ \nu_\gamma $ in the lab frame and photon virtuality $ Q^2 $.

A common way to express the differential inelastic $ e^{-} p $ scattering cross section in terms of the proton structure assumes the exchange of one photon. The cross section is given by the contraction of the leptonic tensor $ L^{\mu \nu} $ and the hadronic tensor $ W^{\mu \nu} $ \cite{Agashe:2014kda}. It is proportional to the phase space of the final lepton (with 4-momentum $ k^\prime = \left( \omega^\prime, \vec{k}^\prime \right)$ in the lab frame) and given by
\ber
d \sigma & = & \frac{e^4}{4 M \sqrt{\omega^2 - m^2} } \frac{\mathrm{d}^3 \vec{k}^\prime} {(2 \pi)^3 2 \omega^\prime}  (4 \pi) L^{\mu \nu}  W_{\mu \nu}, \label{OPE_expressions}
\eer
with the unit of electric charge $ e $. The kinematics are traditionally described by the kinematical Bjorken variable $ x_{\mathrm{Bj}} $, the variable $ y $ related to the energy transferred by the virtual photon relative to the beam energy and the momentum transfer $ Q^2$:
\ber
x_{\mathrm{Bj}} & = & \frac{Q^2}{2(p\cdot q)},  \qquad y = \frac{(p\cdot q)}{(p\cdot k)} = \frac{Q^2}{ 2 x_{\mathrm{Bj}} M \omega}, \qquad Q^2 = - q^2 = - (k-k^\prime)^2. \label{kinematic_variables_xsection}
\eer
The leptonic tensor is evaluated in QED. It is given by
\ber
L^{\mu \nu} & = & 2 \left( k^\mu k^{ \prime \nu} + k^{ \prime \mu} k^{\nu} + ( m^2 - ( k \cdot k^\prime ) ) g^{\mu \nu} - i m \varepsilon^{\mu \nu \rho \sigma} q_\rho s_\sigma \right),
\eer
where $ s^\mu $ is the lepton spin vector:  $ s^\mu s_\mu = - 1, ~ ( s \cdot k ) = 0 $.
The general Lorentz and gauge-invariant structure of the hadronic tensor $ W^{\mu \nu} $ which preserves parity and charge conjugation invariance is given by
\ber \label{inelastic_one_photon}
W_{\mu \nu} & = & \left( - g_{\mu \nu} + \frac{q_\mu q_\nu}{q^2} \right) F_1\left(\nu_\gamma, Q^2\right) + \frac{1}{(p \cdot q)} \left(p^{\mu}-\frac{p\cdot
q}{q^2}\,q^{\mu}\right) \left(p^{\nu}-\frac{p\cdot
q}{q^2}\, q^{\nu} \right)  F_2\left(\nu_\gamma, Q^2\right) \nonumber \\
& + & i \varepsilon_{\mu \nu \alpha \beta} \frac{M q^\alpha}{(p \cdot q)} \left[ S^\beta g_1 \left(\nu_\gamma, Q^2\right) + \left(  S^\beta - \frac{(S \cdot q)}{(p\cdot q)} p^\beta \right) g_2\left(\nu_\gamma, Q^2\right) \right],
\eer
with the virtual photon energy in the laboratory frame $ \nu_\gamma = \left( p \cdot q \right) / M $ and the proton SFs $ F_1\left(\nu_\gamma, Q^2\right) , ~F_2\left(\nu_\gamma, Q^2\right) , ~g_1\left(\nu_\gamma, Q^2\right) , ~g_2\left(\nu_\gamma, Q^2\right)  $, which are extracted from the experimental data. The proton spin 4-vector satisfies: $ S^2 = - 1, ~ (S \cdot p) = 0 $. 

The total unpolarized cross section is given by
\ber \label{fp_cross_section}
\frac{ \mathrm{d}^2 \sigma }{ \mathrm{d} \nu_\gamma d Q^2 } & = & \frac{ \pi \alpha^2}{(Q^2)^2}  \frac{2}{\omega^2 -m^2}  \left(\frac{ Q^2 - 2 m^2}{M} F_1 \left(\nu_\gamma, Q^2\right) + \left( \frac{2 \omega^2}{\nu_\gamma} - 2 \omega  - \frac{Q^2}{2 \nu_\gamma} \right) F_2 \left(\nu_\gamma, Q^2\right) \right).\nonumber \\
\eer
This expression reduces to the well-known expression \cite{Agashe:2014kda} in the massless limit.

Consider a scattering of longitudinally polarized leptons on the proton polarized in the lepton momentum direction $ \sigma_{h \lambda} = \sigma_{\hp \hm} $ and the scattering on the proton polarized in the opposite direction $ \sigma_{h \lambda} = \sigma_{\hp \hp} $ with the proton (lepton) spin vector in the laboratory frame $ S^\mu =  (0, - \lambda \hat{\vec{k}}) $ ($s^\mu = ( |\vec{k}|, \omega \hat{\vec{k}} )$$/m$) and $ \hat{\vec{k}} = \vec{k} / |\vec{k}|$. For the cross sections difference we obtain:
\ber \label{fm_cross_section}
\frac{ \mathrm{d}^2 \sigma_{\hp \hp} - \mathrm{d}^2 \sigma_{\hp \hm}}{ \mathrm{d} \nu_\gamma \mathrm{d} Q^2 } & = &   \frac{4 \pi  \alpha^2}{ \nu_\gamma  M Q^2} \frac{\omega}{\omega^2 - m^2} \left\{ - \frac{Q^2 }{\nu_\gamma \omega} g_2\left(\nu_\gamma, Q^2\right) \right. \nonumber \\
&& \left. + \left( 2 - \frac{Q^2}{2 (\omega^2 - m^2)} \left( 1 + \frac{2 \nu_\gamma m^2}{ Q^2 \omega } \right) \left( 1 + \frac{ 2 \nu_\gamma \omega}{ Q^2 } \right) \right) g_1 \left(\nu_\gamma, Q^2\right) \right\}. 
\eer
This expression reduces to the known expression \cite{Agashe:2014kda}, \cite{Blumlein:1996vs} in the massless limit.

Consider the scattering of transversely polarized leptons on transversely polarized protons. Denoting the averaged over the azimuthal angle cross section $ \sigma_{\perp} $ ($\sigma_{\parallel} $) for scattering with perpendicular (parallel) spin vectors of lepton ($s^\mu = (0, \cos \phi_l, \sin \phi_l, 0)$) and proton ($S^\mu = (0, \cos \phi_p, \sin \phi_p, 0)$), \footnote{The nontrivial relation $ \sigma_{\perp} = \sigma $ holds for the lepton-proton scattering. We have obtained this relation in the OPE approximation of Eq. (\ref{OPE_expressions}) and have proved it by the direct cross sections evaluation for the case of elastic scattering and exploiting the symmetry properties of the helicity amplitudes for the scattering to arbitrary channel $ lp \to l X $.} i. e. $ \phi_l - \phi_p = \pm \pi /2 $ ($\phi_l - \phi_p = 0 $) for the perpendicular (parallel) configuration, we obtain:
\ber \label{g_cross_section}
\frac{ \mathrm{d}^2 \sigma_{\perp}  - \mathrm{d}^2 \sigma_{\parallel} }{ \mathrm{d} \nu_\gamma \mathrm{d} Q^2 } & = & \frac{2 \pi m \alpha^2}{ \nu_\gamma  M Q^2} \frac{1}{\omega^2 - m^2} \left\{  2 g_2 \left(\nu_\gamma, Q^2\right)  \right. \nonumber \\
&& \left. +  \left( 1 + \frac{ \omega \nu_\gamma}{\omega^2 -m^2} \left( 1 + \frac{m^2 \nu_\gamma}{\omega Q^2}   + \frac{Q^2}{4 \nu_\gamma \omega} \right) \right) g_1 \left(\nu_\gamma, Q^2\right) \right\}.
\eer
Consequently, a measurement of the inclusive $e^- p$ cross sections accesses the proton spin SFs $ g_1 $ and $ g_2 $. 

The elastic scattering cross sections $ l^- p \to l^- p $ are obtained by substitution of the inelastic SFs in Eqs. (\ref{fp_cross_section}-\ref{g_cross_section}) by the elastic contribution to them:
\ber
F_1 \left(x_{\mathrm{Bj}},Q^2\right) & \to & \frac{1}{2} G^2_M \left(Q^2\right) \delta \left( 1 - x_{\mathrm{Bj}} \right), \\
F_2 \left(x_{\mathrm{Bj}},Q^2\right) & \to & \frac{G^2_E \left(Q^2\right) + \tau_P G^2_M \left(Q^2\right)}{1 + \tau_P}  \delta \left( 1 - x_{\mathrm{Bj}} \right),\\
g_1 \left(x_{\mathrm{Bj}},Q^2\right) & \to &  \frac{1}{2}  F_D \left(Q^2\right) G_M \left(Q^2\right) \delta \left(1 - x_{\mathrm{Bj}} \right), \\
g_2 \left(x_{\mathrm{Bj}},Q^2\right) & \to & -  \frac{1}{2}  \tau_P F_P \left(Q^2\right) G_M \left(Q^2\right) \delta \left(1 - x_{\mathrm{Bj}} \right), \label{g2_ampl}
\eer
where $ F_D (Q^2) $, $ F_P(Q^2) $, $ G_E (Q^2) $, $ G_M(Q^2) $ are the Dirac, Pauli, Sachs electric and magnetic proton form factors and $ \tau_P = Q^2/( 4 M^2 ) $, following by the integration over the virtual photon energy.

Substituting the expressions for the inclusive cross sections of Eqs. (\ref{fp_cross_section}-\ref{g_cross_section}) into the DRs, see Eqs. (\ref{M2_DR}-\ref{M3_DR}), changing the integration order, as detailed in Appendix \ref{relations_between_amplitudes}, and expressing the spin-dependent forward TPE amplitudes in terms of the proton SFs, we obtain:
\ber
\Re g^{2 \gamma} \left(\omega\right) & = & \frac{4 m \alpha^2}{\vec{k}^2} \int \limits^{\infty}_{0} \frac{\mathrm{d} Q^2}{Q^2} \int \limits^{\infty}_{\nu_{\mathrm{thr}}} \frac{\mathrm{d} \nu_\gamma}{\nu_\gamma} \nonumber \\
&& \left \{ 2 \frac{ (\omega_0 - |\vec{k}_0| ) \nu_\gamma + m^2 (\tau_l + \tilde{\tau})}{|\vec{k}_0|} g_1 \left( \nu_\gamma, Q^2 \right) \right. \nonumber \\
&& \left. + \frac{ m^2 (\tau_l + \tilde{\tau}) + \vec{k}^2 }{|\vec{k}|} g_1 \left( \nu_\gamma, Q^2 \right) \ln \frac{|\vec{k}|-|\vec{k}_0|}{|\vec{k}|+|\vec{k}_0|} + 2 g_2 \left( \nu_\gamma, Q^2 \right) \ln \frac{|\vec{k}|-|\vec{k}_0|}{|\vec{k}|+|\vec{k}_0|} \right. \nonumber \\
&& \left. + \frac{\omega \nu_\gamma }{|\vec{k}|} g_1 \left( \nu_\gamma, Q^2 \right) \ln \frac{ \left( \omega+|\vec{k}| \right)^2 \left(\omega_0^2 - \omega^2 \right)}{ \left( \omega |\vec{k}_0| + |\vec{k}| \omega_0 \right)^2 } \right \},  \label{M3} \\
\Re f^{2 \gamma}_- \left(\omega\right) & = & \frac{8 \omega \alpha^2}{\vec{k}^2} \int \limits^{\infty}_{0} \frac{\mathrm{d} Q^2}{Q^2} \int \limits^{\infty}_{\nu_{\mathrm{thr}}} \frac{\mathrm{d} \nu_\gamma}{\nu_\gamma} \nonumber \\
&& \left \{  2 \frac{ (\omega_0 - |\vec{k}_0| ) \nu_\gamma + m^2 (\tau_l + \tilde{\tau})}{ |\vec{k}_0|} g_1 \left( \nu_\gamma, Q^2 \right)   \right. \nonumber \\
&& \left. + \frac{  m^2 (\tau_l + \tilde{\tau}) - \vec{k}^2 }{|\vec{k}|} g_1 \left( \nu_\gamma, Q^2 \right) \ln \frac{|\vec{k}|-|\vec{k}_0|}{|\vec{k}|+|\vec{k}_0|} \right. \nonumber \\
&& \left. + \frac{Q^2 |\vec{k}| }{ 2 \omega \nu_\gamma} g_2 \left( \nu_\gamma, Q^2 \right) \ln \frac{ \left( \omega+|\vec{k}| \right)^2 \left(\omega_0^2 - \omega^2 \right)}{ \left( \omega |\vec{k}_0| + |\vec{k}| \omega_0 \right)^2 } \right. \nonumber \\
&& \left. + \frac{(\omega^2 + m^2) \nu_\gamma }{2 \omega |\vec{k}|} g_1 \left( \nu_\gamma, Q^2 \right) \ln \frac{ \left( \omega+|\vec{k}| \right)^2 \left(\omega_0^2 - \omega^2 \right)}{ \left( \omega |\vec{k}_0| + |\vec{k}| \omega_0 \right)^2 } \right \},  \label{M2}
\eer
where we have introduced the notations:
\ber
|\vec{k}| & = & \sqrt{\omega^2 - m^2}, \quad |\vec{k}_0| = \sqrt{ \omega_0^2 - m^2 }, \\
\omega_0 & = & m \left( \sqrt{\tau_l\tilde{\tau}} +  \sqrt{1+\tau_l} \sqrt{1+\tilde{\tau}}\right), \\
 \tau_l & = & \frac{Q^2}{4 m^2},~~~~  \tilde{\tau} = \frac{\nu_\gamma^2}{Q^2}. \label{taus}
\eer
In Eqs. (\ref{M2}, \ref{M3}) the elastic threshold $ \nu_{\mathrm{thr}} $ and  the inelastic threshold $ \nu^{\mathrm{inel}}_{\mathrm{thr}} $ are given by $ \nu_{\mathrm{thr}} = 0 $ and $ \nu^{\mathrm{inel}}_{\mathrm{thr}} = m_\pi +  \left( m_\pi^2 + Q^2 \right) / \left( 2 M \right)$ respectively, where $ m_{\pi} $ denotes the pion mass.

The leading TPE correction to the atomic energy levels is given by the values of the amplitudes at threshold $ \omega = m $. The TPE amplitudes $ f_-^{2\gamma}, ~g^{2\gamma}$ at threshold can then be expressed in terms of the proton spin SFs as

\ber
f_-^{2\gamma} \left( m \right ) & = & \frac{16 \alpha^2}{3} \int \limits^{\infty}_{0} \frac{\mathrm{d} Q^2}{Q^2} \int \limits^{\infty}_{\nu_{\mathrm{thr}}} \frac{\mathrm{d} \nu_\gamma}{\nu_\gamma}  \frac{ \left [ 2 + \rho\left(\tau_l\right) \rho\left(\tilde{\tau}\right) \right ]  g_1 \left(\nu_\gamma, Q^2 \right)  -  3 \rho\left(\tau_l\right) \rho\left(\tilde{\tau} \right)  g_2 \left(\nu_\gamma, Q^2 \right)   / \tilde{\tau}}{  \sqrt{\tilde{\tau}} \sqrt{1+\tau_l} + \sqrt{\tau_l} \sqrt{1+\tilde{\tau}}   } , \nonumber \\ \label{fminus}  \\
g^{2\gamma} \left( m \right ) & = & - \frac{16 \alpha^2}{3} \int \limits^{\infty}_{0} \frac{\mathrm{d} Q^2}{Q^2} \int \limits^{\infty}_{\nu_{\mathrm{thr}}} \frac{\mathrm{d} \nu_\gamma}{\nu_\gamma}  \frac{  \left [ 2 + \rho\left(\tau_l\right) \rho\left(\tilde{\tau}\right) \right ]  g_1 \left(\nu_\gamma, Q^2 \right) + 3  g_2 \left(\nu_\gamma, Q^2 \right) }{  \sqrt{\tilde{\tau}} \sqrt{1+\tau_l} + \sqrt{\tau_l} \sqrt{1+\tilde{\tau}}   }, \label{g_ampl}
\eer
with
\ber
\rho(\tau) & = &\tau - \sqrt{\tau ( 1 + \tau )}. \label{rho_notation}
\eer

Evaluating the sum of the spin-dependent lepton-proton TPE amplitudes in the DR approach, see Eqs. (\ref{fminus}) and (\ref{g_ampl}), we obtain:
\ber \label{Old_result_condition}
g^{2\gamma} (m) + f_-^{2\gamma}(m) =  64 \alpha^2 M m \int \limits^{\infty}_{0} \frac{\mathrm{d} Q^2}{Q^4} \rho \left( \tau_l \right) \int \limits^{1}_0 \mathrm{d} x_\mathrm{Bj} g_2 \left( x_\mathrm{Bj}, Q^2 \right) = 0,
\eer
which is a trivial relation due to the Burkhardt-Cottingham (BC) sum rule \cite{Burkhardt:1970ti}:
\ber
\int \limits^{1}_0 \mathrm{d} x_\mathrm{Bj} g_2 \left( x_\mathrm{Bj}, Q^2 \right) = 0.
\eer

\section{Lamb shift and hyperfine splitting (HFS)}
\label{hfs_correction}

In this Section, we express the TPE correction to S energy levels in the hydrogen-like atom in terms of the forward TPE amplitudes at threshold and compare our result with the standard approach.

In the $ lp $ center-of-mass (c.m.) reference frame the TPE forward scattering amplitude $ T^{2\gamma} $ is expressed in terms of the forward amplitudes $ f^{2\gamma}_+,~f^{2\gamma}_-,~g^{2\gamma} $ as
\ber
T^{2\gamma}(\omega) & = & f^{2\gamma}_+ \left(\omega\right) +  4 g^{2\gamma}\left(\omega\right) \, \boldsymbol{s} \cdot \boldsymbol{S} 
+  4 \big(f^{2\gamma}_-\left(\omega\right) + g^{2\gamma}\left(\omega\right) \big) \, \boldsymbol{s}  \cdot \hat{\vec{k}}\, \boldsymbol{S}  \cdot \hat{\vec{p}},
\eer
with $ \boldsymbol{s} $ ($ \boldsymbol{S} $) and $ \hat{\vec{k}} $ ($ \hat{\vec{p}} $) the lepton (proton) spin and momentum direction vectors. This decomposition often arises in the analysis of the non-relativistic forward neutron-proton scattering, see e.g.  \cite{Guichon:1982eb}.

It is then easy to see that, considering $ T^{2\gamma} $ as correction to the Coulomb potential, its effect on the nS-state energy level is given by
\ber  \label{2S_correction_Lamb_shift}
\Delta E_{\mathrm{nS}} = - \frac{|\psi_{\mathrm{nS}}\left(0\right)|^2}{4 M m} f^{2\gamma}_+\left( m \right),
\eer
with $ | \psi_{\mathrm{nS}} (0) |^2 =  \alpha^3 m_r^3/(\pi n^3) $ - the non-relativistic squared wave function of the hydrogen atom, where $ m_r = M m / ( M + m )$ is the reduced mass of the lepton and proton bound state. Using the DR for $f_+^{2\gamma}$ with only the elastic part of the unpolarized cross section and subtracting the accounted TPE contribution in the hydrogen wave functions, as well as the OPE finite-size correction, we reproduce the non-relativistic limit of the TPE contribution:
\ber
\Delta E_{\mathrm{nS}} = - \frac{8 m_r^4 \alpha^5}{\pi n^3} \int \limits^{\infty}_{0} \frac{\mathrm{d} Q^2}{Q^5} \left( G^2_E(Q^2) - 2 G_E' (0) Q^2 - 1 \right). \label{Lamb_shift_non_rel}
\eer
This correction yields the third Zemach moment \cite{Hagelstein:2015egb}.

The TPE contribution to the nS-level HFS $ \delta E^{\mathrm{HFS}}_{\mathrm{nS}} $ is expressed in terms of the relative correction $ \Delta_{\mathrm{HFS}} $ and the leading order nS-level HFS $ E^{\mathrm{HFS},0}_{\mathrm{nS}} $ (Fermi energy) as \footnote{Note the absence of the muon anomalous magnetic moment in Eq. (\ref{fermi_energy}) \cite{Eides:2000xc}.} 
\ber
 \delta E^{\mathrm{HFS}}_{\mathrm{nS}} & = & \Delta_{\mathrm{HFS}} E^{\mathrm{HFS},0}_{\mathrm{nS}} , \\
 E^{\mathrm{HFS},0}_{\mathrm{nS}} & = & \frac{8}{3} \frac{ m_r^3 \alpha^4}{ M m} \frac{\mu_P}{n^3}. \label{fermi_energy}
\eer

Considering the spin part of $ T^{2\gamma} $ as correction to the Hamiltonian of the lepton-proton spin-spin interaction, we express the leading TPE proton structure correction to the S-level HFS in terms of the amplitudes $ f_-^{2\gamma},~g^{2\gamma} $ at threshold ($ \omega = m $):
\ber \label{2S_correctionx}
\mu_P e^2 \Delta_{\mathrm{HFS}} & = & - g^{2\gamma} \left( m \right) + \frac{1}{2} f_-^{2\gamma}\left( m \right).
\eer

The TPE correction to the S-level HFS $ \Delta_{\mathrm{0}} $ of Refs. \cite{Iddings:1959zz,Iddings:1965zz,Drell:1966kk,Faustov:1966,Faustov:1970,Bodwin:1987mj,Faustov:2001pn,Carlson:2008ke,Carlson:2011af,Peset:2016wjq}, which was derived from the difference between the spin-averaged forward matrix elements in the singlet and triplet states with a subsequent loop integration, can be also obtained adding $  g^{2\gamma}(m) + f_-^{2\gamma}(m) = 0 $ to Eq. (\ref{2S_correctionx}):
\ber \label{2S_correction_oldxx}
\mu_P e^2 \Delta_{\mathrm{0}} & = & \frac{3}{2} f_-^{2\gamma}\left( m \right).
\eer
Consequently, we have verified the TPE correction to HFS of S energy levels: $ \Delta_{\mathrm{0}} = \Delta_{\mathrm{HFS}}$. The HFS correction of Refs. \cite{Iddings:1959zz,Iddings:1965zz,Drell:1966kk,Faustov:1966,Faustov:1970,Bodwin:1987mj,Faustov:2001pn,Carlson:2008ke,Carlson:2011af} can be reproduced performing the loop integration for the amplitudes $ g^{2\gamma}(m)$ and $f_-^{2\gamma}(m)$ expressions in terms of the forward Compton scattering as a lower part of the TPE graph, see Appendix \ref{box_graph_method} for details.

In the following, we study the difference in the individual channel contribution to HFS correction between the traditional HFS expressions and the DR approach based on the unitarity and analyticity of the forward lepton-proton scattering amplitudes. Such a difference between the direct box graph evaluation and unsubtracted dispersion relation result was studied in Refs. \cite{Borisyuk:2008es,Tomalak_PhD,Blunden:2017nby} in the case of the TPE correction to the non-forward elastic lepton-proton scattering.

Traditionally, the TPE correction to HFS is expressed by a sum of the proton state contribution $\Delta^{\mathrm{el}}_{0}$ and the polarizability correction $ \Delta^{\mathrm{pol}}_{\mathrm{0}} $:
\ber
 \Delta_{\mathrm{0}} = \Delta^{\mathrm{el}}_{0} + \Delta^{\mathrm{pol}}_{\mathrm{0}}.
\eer

The proton intermediate state TPE correction to HFS $ \Delta^{\mathrm{el}}_{0}  $ is expressed as a sum of the Zemach correction $ \Delta_{\mathrm{Z}} $ with subtraction of the TPE contribution, which is already accounted for in the hydrogen wave functions, and the recoil correction $ \Delta^{\mathrm{p}}_{\mathrm{R}} $:

\ber \label{traditional}
\Delta^{\mathrm{el}}_{0}  & = & \Delta_{\mathrm{Z}} + \Delta^{\mathrm{p}}_{\mathrm{R}},
\label{Zemach_correction} \\
\Delta_{\mathrm{Z}}& = &\frac{8 \alpha m_r}{\pi \mu_P}  \int \limits^{\infty}_{0} \frac{\mathrm{d} Q}{Q^2} \left( G_M\left(Q^2\right)  G_E\left(Q^2\right)  - \mu_P \right),\\ \Delta^{\mathrm{p}}_{\mathrm{R}}  & = &  \frac{\alpha}{\pi \mu_P} \int \limits^{\infty}_{0} \frac{\mathrm{d} Q^2}{Q^2} \left\{ \frac{  \left [ 2 + \rho\left(\tau_l\right) \rho\left(\tau_P \right) \right ]  F_D\left(Q^2\right) + 3 \rho\left(\tau_l\right) \rho \left(\tau_P\right) F_P\left(Q^2\right)   }{ \sqrt{\tau_P} \sqrt{1+\tau_l} + \sqrt{\tau_l} \sqrt{1+\tau_P} } - \frac{4 m_r}{Q} G_E\left(Q^2\right)  \right\} \nonumber \\ 
&& \qquad \qquad \quad ~~ \times G_M\left(Q^2\right) - \frac{\alpha}{\pi \mu_P} \frac{m}{M} \int \limits^{\infty}_{0} \frac{\mathrm{d} Q}{Q} \beta_1(\tau_l) F_P^2 \left(Q^2\right), \label{recoil_correction}
\eer
with $ \beta_1(\tau) = -3 \tau + 2 \tau^2 +2 (2 - \tau) \sqrt{\tau (1+\tau)}$ \cite{Carlson:2008ke}.
The substitution of the photon-proton-proton vertex in the TPE box graph by the on-shell vertex with the Dirac and Pauli couplings \cite{Iddings:1959zz,Iddings:1965zz,Blunden:2003sp}, which reproduces the proton state contribution to the VVCS tensor, with a subsequent loop integration, leads to the correction $ \Delta^{\mathrm{el}}_{0}$, while the result based on the dispersion relations for the VVCS tensor $  \Delta^{\mathrm{p}}_{0}$ is given by
\ber
  \Delta^{\mathrm{p}}_{0} =  \Delta^{\mathrm{el}}_{0}  +  \Delta^{\mathrm{F^2_P}},
\eer
which differs from $ \Delta^{\mathrm{el}}_{0} $ by the non-pole contribution $ \Delta^{\mathrm{F^2_P}} $:
\ber \label{non_pole}
 \Delta^{\mathrm{F^2_P}} = \frac{\alpha}{\pi \mu_P} \frac{m}{M} \int \limits^{\infty}_{0} \frac{\mathrm{d} Q}{Q} \beta_1(\tau_l) F_P^2 \left(Q^2\right).
\eer
This correction comes from the proton non-pole term in the spin-dependent forward double virtual Compton scattering tensor, see Appendix \ref{box_graph_method} for a description of the VVCS tensor, and should not contribute to the resulting TPE correction as the non-pole terms are not reproduced by the dispersion relations \cite{Carlson:2011af}.

We express the elastic TPE contribution to the S-level HFS $ \Delta^{\mathrm{el}}_{\mathrm{HFS}} $ in the lepton-proton amplitudes DR framework with subtraction of the TPE contribution, which is already accounted for in the hydrogen wave functions, in terms of the proton electric and magnetic form factors as 
\ber
 \Delta^{\mathrm{el}}_{\mathrm{HFS}}  & = & \frac{\alpha}{\pi \mu_P} \int \limits^{\infty}_{0} \frac{\mathrm{d} Q^2}{Q^2} \left\{ \frac{ 2 G_E\left(Q^2\right) +  \rho \left(\tau_l\right) \rho\left(\tau_P\right)   G_M\left(Q^2\right)  }{ \sqrt{\tau_P} \sqrt{1+\tau_l} + \sqrt{\tau_l} \sqrt{1+\tau_P} } G_M\left(Q^2\right)-  \frac{4 \mu_P m_r}{Q} \right\}. \label{elastic_correction}
\eer

We reproduce the Zemach correction \cite{Zemach:1956zz} as the non-relativistic limit of the elastic TPE correction $ \Delta^{\mathrm{el}}_{\mathrm{HFS}} $. In our approach, the non-pole term of Eq. (\ref{non_pole}) does not appear. The difference between the proton state contribution in the box graph model with an on-shell vertex \cite{Iddings:1959zz,Iddings:1965zz,Blunden:2003sp} $ \Delta^{\mathrm{p}}_{0} $ and within dispersion relation approach $ \Delta^{\mathrm{el}}_{\mathrm{HFS}}  $, see Eqs. (\ref{traditional}-\ref{non_pole}) and Eq. (\ref{elastic_correction}), $\Delta^{\mathrm{p}}_{0}  - \Delta^{\mathrm{el}}_{\mathrm{HFS}}  $ is given by the elastic contribution to the amplitude $ g^{2\gamma}(m) + f_-^{2\gamma}(m)$.

Traditionally, the polarizability correction $ \Delta^{\mathrm{pol}}_{\mathrm{0}} $ is given by \cite{Carlson:2008ke,Carlson:2011af}
\ber
\Delta^{\mathrm{pol}}_{\mathrm{0}}   =   \frac{2 \alpha}{\pi \mu_P} \int \limits^{\infty}_{0} \frac{\mathrm{d} Q^2}{Q^2} \int \limits^{\infty}_{\nu^{\mathrm{inel}}_{\mathrm{thr}}} \frac{\mathrm{d} \nu_\gamma}{\nu_\gamma}  \frac{\left [ 2 + \rho\left(\tau_l\right) \rho\left(\tilde{\tau}\right) \right ]  g_1 \left(\nu_\gamma, Q^2 \right)  -  3 \rho\left(\tau_l\right) \rho\left(\tilde{\tau} \right)  g_2 \left(\nu_\gamma, Q^2 \right)   / \tilde{\tau}}{  \sqrt{\tilde{\tau}} \sqrt{1+\tau_l} + \sqrt{\tau_l} \sqrt{1+\tilde{\tau}}   }  +  \Delta^{\mathrm{F^2_P}}. \nonumber \\
\eer
The non-pole term $  \Delta^{\mathrm{F^2_P}} $ allows one to expand the HFS integrand near $ Q^2 =0 $ in terms of the proton spin polarizabilities \cite{Hagelstein:2015egb} and to be consistent with the massless lepton limit of this correction \cite{Carlson:2011af}. In order to avoid the non-pole contributions in the resulting TPE correction to HFS, it was subtracted from the recoil correction in Refs. \cite{Carlson:2008ke,Carlson:2011af}, see Eq. (\ref{recoil_correction}).

We express the inelastic $\alpha^5$-correction to the S-level HFS in the lepton-proton amplitudes DR approach in terms of the proton inelastic spin SFs $ g_1 $ and $ g_2 $ as
\ber
\Delta^{\mathrm{inel}}_{\mathrm{HFS}}  & = & \frac{2 \alpha}{\pi \mu_P} \int \limits^{\infty}_{0} \frac{\mathrm{d} Q^2}{Q^2} \int \limits^{\infty}_{\nu^{\mathrm{inel}}_{\mathrm{thr}}} \frac{\mathrm{d} \nu_\gamma}{\nu_\gamma}   \frac{\left [ 2 + \rho\left(\tau_l\right) \rho\left(\tilde{\tau}\right) \right ]  g_1 \left(\nu_\gamma, Q^2 \right) + \left [ 2 - \rho\left(\tau_l\right) \rho\left(\tilde{\tau}\right)/ \tilde{\tau}  \right ]  g_2 \left(\nu_\gamma, Q^2 \right)   }{   \sqrt{\tilde{\tau}} \sqrt{1+\tau_l} + \sqrt{\tau_l} \sqrt{1+\tilde{\tau}}   }. \nonumber \\
 \label{inelastic_correct}
\eer
The remaining difference in the inelastic correction $ \Delta^{\mathrm{pol}}_{\mathrm{0}}  - \Delta^{\mathrm{F^2_P}} -  \Delta^{\mathrm{inel}}_{\mathrm{HFS}} $ is given by the contribution from the spin SF $ g_2 $ to the amplitude $ g^{2\gamma}(m) + f_-^{2\gamma}(m) $.

Consequently, the resulting TPE correction to HFS of S energy levels can be equivalently expressed as
\ber
\Delta_{\mathrm{HFS}} = \Delta^{\mathrm{el}}_{\mathrm{HFS}} + \Delta^{\mathrm{inel}}_{\mathrm{HFS}}  = \Delta_{\mathrm{Z}} + \Delta^{\mathrm{p}}_{\mathrm{R}} + \Delta^{\mathrm{pol}}_{\mathrm{0}} =  \Delta_0.
\eer

\section{HFS correction in both approaches}
\label{hfs_correction_numbers}

For the numerical evaluation of the TPE corrections to HFS from the proton intermediate state and the $  \Delta^{\mathrm{F^2_P}} $ part of the polarizability correction $ \Delta^{\mathrm{pol}}_{\mathrm{0}} $ we exploit the elastic form factor parametrizations from Refs. \cite{Bernauer:2010wm,Bernauer:2013tpr}. For the Zemach correction we make two evaluations for the 1-$\sigma$ band curves coming from the elastic proton form factor uncertainties of Refs. \cite{Bernauer:2010wm,Bernauer:2013tpr}, where a global analysis of the electron-proton scattering data with account of TPE corrections for $ Q^2 < 10 ~\mathrm{GeV}^2 $ was performed, and estimate the uncertainty as a half of a difference between these two curves. For the numerical evaluation of the inelastic correction we exploit the spin SFs data parametrization from Refs. \cite{Kuhn:2008sy,griffioen,Sato:2016tuz} in the region of large $ Q^2 $. In the region of low $Q^2$, we expand the $ Q^2 $-integrand from the proton spin SFs in terms of small $ x_{\mathrm{Bj}} $ and account for the leading non-vanishing moments:
\ber
\Delta^{\mathrm{pol}}_0 & \to & \frac{\alpha}{2 \pi } \int \limits_0^{} \mathrm{d} Q^2 \frac{\rho \left( \tau_l \right) \left( \rho \left( \tau_l \right) - 4 \right) }{\mu_P M m \tau_l} I_1 \left( Q^2 \right) + \frac{\alpha}{8 \pi} \int \limits_0^{} \mathrm{d} Q^2 \frac{ \left( 9 - 2 \rho \left( \tau_l \right) \right) \rho \left( \tau_l \right)^2 }{\mu_P M m \tau_l} I^{(3)}_1 \left( Q^2 \right)  \nonumber \\
& - &  \frac{3 \alpha}{ 2 \pi } \int \limits_0^{} \mathrm{d} Q^2 \frac{ 1 + 2 \rho \left( \tau_l \right) }{\mu_P M m} I^{(3)}_2 \left( Q^2 \right), \\
\Delta^{\mathrm{inel}}_{\mathrm{HFS}}  & \to & \Delta^{\mathrm{pol}}_0 - \frac{2 \alpha }{\pi} \int \limits_0^{} \mathrm{d} Q^2  \frac{\rho \left( \tau_l \right) }{\mu_P M m \tau_l} I_2 \left( Q^2 \right),
\eer
with the moments of the proton spin SFs:
\ber 
I_1 \left( Q^2 \right) & = & \frac{2M^2}{Q^2} \int \limits^{x^\mathrm{inel}_\mathrm{thr}}_0 g_1 \left( x_{\mathrm{Bj}}, ~Q^2\right) \mathrm{d} x_{\mathrm{Bj}}, \qquad \quad I_1(0) = - \frac{\left(\mu_P - 1 \right)^2}{4}, \label{I1_moment} \\
I_2 \left( Q^2 \right) & = & \frac{2M^2}{Q^2} \int \limits^{x^\mathrm{inel}_\mathrm{thr}}_0 g_2 \left( x_{\mathrm{Bj}}, ~Q^2\right) \mathrm{d} x_{\mathrm{Bj}} = \frac{1}{4} F_P \left( Q^2 \right) G_M \left( Q^2 \right), \label{I2_moment} \\
I^{(3)}_1 \left( Q^2 \right) & = & \frac{8M^4}{Q^4} \int \limits^{x^\mathrm{inel}_\mathrm{thr}}_0 x_{\mathrm{Bj}}^2 g_1 \left( x_{\mathrm{Bj}}, ~Q^2\right) \mathrm{d} x_{\mathrm{Bj}} \underset{Q^2 \to 0}{\longrightarrow} \frac{Q^2 M^2}{2 \alpha} \gamma_0,  \\
I^{(3)}_2 \left( Q^2 \right) & = & \frac{8M^4}{Q^4} \int \limits^{x^\mathrm{inel}_\mathrm{thr}}_0 x_{\mathrm{Bj}}^2 g_2 \left( x_{\mathrm{Bj}}, ~Q^2\right) \mathrm{d} x_{\mathrm{Bj}} \underset{Q^2 \to 0}{\longrightarrow} \frac{Q^2 M^2}{2 \alpha} \left( \delta_{\mathrm{LT}} - \gamma_0 \right), \label{I32_moment}
\eer
with $ x^\mathrm{inel}_\mathrm{thr} = Q^2/ (2 M \nu^{\mathrm{inel}}_{\mathrm{thr}}) $ and the low-energy constants values \cite{Drechsel:2000ct,Drechsel:2007if,Drechsel:2002ar,Prok:2008ev,Pascalutsa:2014zna,Lensky:2017dlc}:
\ber
 \delta_{\mathrm{LT}} & = & \left( 1.34 \pm 0.17 \right) \times 10^{-4} ~\mathrm{fm}^4, \\
  \gamma_0 & = & \left( -1.01 \pm 0.13\right) \times 10^{-4} ~\mathrm{fm}^4, \\
{I'}_1 \left( 0 \right) & = & \left( 7.6 \pm 2.5 \right) ~\mathrm{GeV}^{-2}.
\eer

\begin{figure}[H]
\begin{center}
\includegraphics[width=0.4\textwidth]{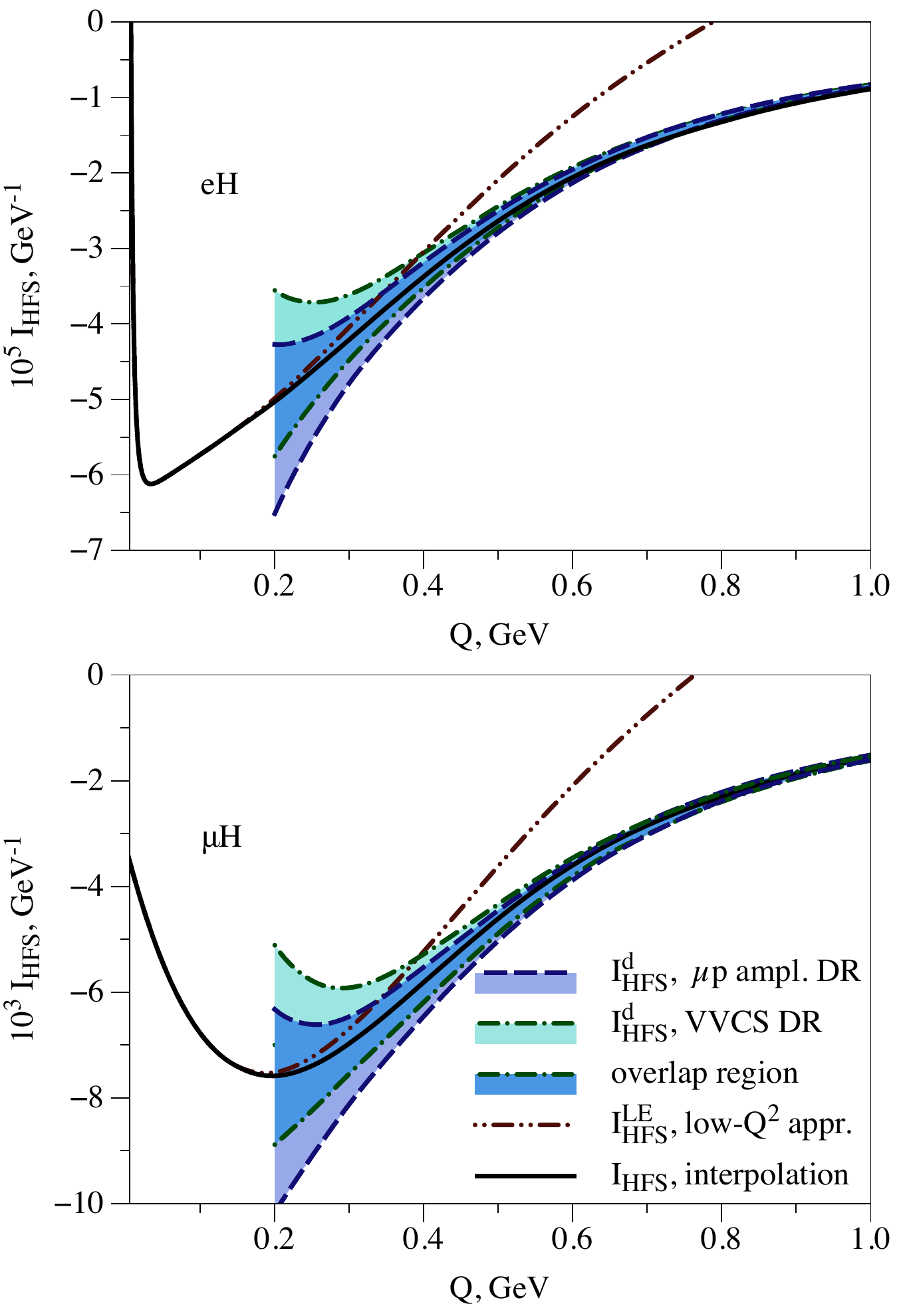}
\end{center}
\caption{Q-dependence of the integrand $ I_{\mathrm{HFS}} \left( Q \right) $ entering the TPE correction to HFS. A comparison is given for the integrands based on the DRs for the $lp$ amplitudes and based on the forward double virtual Compton scattering (VVCS) amplitudes. Upper panel: electronic hydrogen, lower panel: muonic hydrogen.}
\label{HFS_integrand}
\end{figure}

In Fig. \ref{HFS_integrand} we show the integrand $I_{\mathrm{HFS}} (Q)$ entering the TPE correction:
\ber \label{HFS_Q_integrand}
\Delta_{\mathrm{HFS}} = \int \limits^{\infty}_0 I_{\mathrm{HFS}}(Q) \mathrm{d} Q,
\eer
\noindent in the case of $ e \mathrm{H} $ and $ \mu \mathrm{H} $. The low-$ Q $ behavior based on the moments of the proton spin SFs of Eqs. (\ref{I1_moment}, \ref{I2_moment}-\ref{I32_moment}) $I^{\mathrm{LE}}_{\mathrm{HFS}}$ and the high-$ Q $ behavior based on the data $I^{\mathrm{d}}_{\mathrm{HFS}}$ are almost independent of the way to evaluate the TPE correction (for $ Q > 0.5 ~\mathrm{GeV} $ the two methods agree within $ 2.5 \% $). While in the region $ 0.2~\mathrm{GeV} \lesssim Q \lesssim 0.5~\mathrm{GeV} $ the HFS evaluation with the DRs for the lepton-proton amplitudes ($  \Delta_{\mathrm{HFS}} = \Delta_{\mathrm{HFS}}^{\mathrm{el}} + \Delta_{\mathrm{HFS}}^{\mathrm{inel}} $) and the traditional HFS evaluation ($  \Delta_{\mathrm{HFS}} = \Delta_{0} = \Delta^{\mathrm{el}}_{0}   + \Delta_0^{\mathrm{pol}} $) slightly differ. New data in this kinematical region will be very useful for such evaluation. In order to avoid any model dependence, we connect the two model-independent regions by the function of the Fermi-Dirac distribution type:
\ber
I_{\mathrm{HFS}}(Q) & = & I^{\mathrm{LE}}_{\mathrm{HFS}} \left( Q \right) \Theta \left(Q_{\mathrm{LE}} - Q \right) + I^{\mathrm{d}}_{\mathrm{HFS}} \left( Q \right) \Theta \left( Q - Q_{\mathrm{d}} \right) + \nonumber \\ 
&& \frac{c_1 I^{\mathrm{LE}}_{\mathrm{HFS}} \left( Q \right) +c_2 f \left(Q\right) I^{\mathrm{d}}_{\mathrm{HFS}} \left( Q \right) }{1+f \left(Q \right)} \Theta \left( Q -Q_{\mathrm{LE}}\right) \Theta \left( Q_{\mathrm{d}} -Q \right),
\eer
with $ f(Q) $ given by
\ber
f \left(Q\right) = e^\frac{2 Q - Q_{\mathrm{LE}} - Q_{\mathrm{d}}}{2 a_0}.
\eer
Furthermore, $ Q_{\mathrm{LE}} = 0.2 ~\mathrm{GeV} $, $ Q_{\mathrm{d}} = 0.5 ~\mathrm{GeV} $, $ a_0 = 0.1~\mathrm{GeV} $, and the constants $ c_1, ~c_2,~Q_{\mathrm{LE}}, ~a_0$ were chosen as those that preserve the regularity and smoothness of the integrand $ I_{\mathrm{HFS}}(Q) $.

\begin{figure}[H]
\begin{center}
\includegraphics[width=0.54\textwidth]{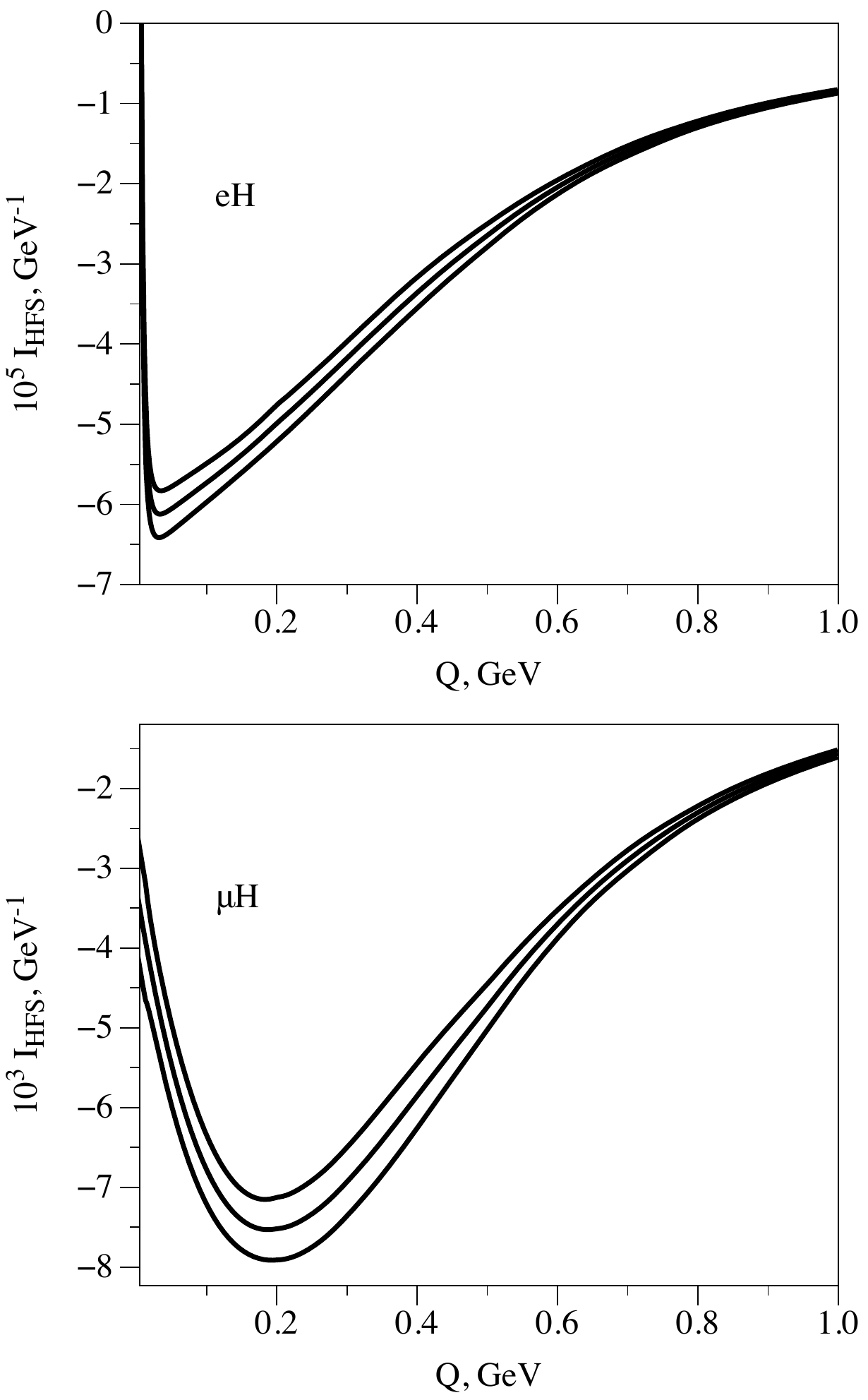}
\end{center}
\caption{HFS integrand $ I_{\mathrm{HFS}} \left( Q \right) $ in the evaluation by the DRs for $lp$ amplitudes with error bands. Upper panel: electronic hydrogen, lower panel: muonic hydrogen.}
\label{HFS_integrand_errors}
\end{figure}

In the low-$Q$ region, we make two evaluations for the 1-$\sigma$ bands of the elastic proton form factors from Refs. \cite{Bernauer:2010wm,Bernauer:2013tpr}. We add the combined uncertainty from $ \gamma_0, ~\delta_{LT}, ~{I'}_1 \left(0\right)$ linearly to the form factors uncertainty. For the larger $ Q > (0.013-0.017) ~\mathrm{GeV} $, we make evaluation for the central values of the proton elastic form factors and add the uncertainty of the proton elastic form factors to the uncertainties from $ \gamma_0, ~\delta_{LT}, ~{I'}_1 \left(0\right)$ in quadrature. The boundary $ Q$ value is chosen as the value that leads to the same uncertainties in the proton intermediate state HFS contribution in both ways of the error estimate described in this paragraph. For the larger $ Q > 0.5 ~\mathrm{GeV} $ region,  we add the uncertainty of the proton spin structure function parametrization in quadrature to the uncertainties coming from the proton elastic form factors. We connect the high-$Q$ integrands $ I^{\mathrm{d}} $ by two curves to the 1-$\sigma$ boundaries in the low-$ Q $ region $ I^{\mathrm{LE}} $. We estimate the uncertainty from the difference between the integral of Eq. (\ref{HFS_Q_integrand}) for these two curves, which are shown in Fig. \ref{HFS_integrand_errors} for $ e \mathrm{H} $ ($ \mu \mathrm{H} $), and take the averaged central value. In the region $ Q^2 > 10 ~\mathrm{GeV}^2 $, the sizable contribution comes only from the $\mu_P $ term in Eqs. (\ref{Zemach_correction}-\ref{elastic_correction}) and does not introduce any sizable additional uncertainty.

We evaluate the proton TPE correction to HFS in the DR approach either for the forward double virtual Compton scattering (TPE correction of Refs. \cite{Carlson:2008ke,Carlson:2011af}) or for the lepton-proton amplitudes. Also we present results for the Zemach correction, the recoil correction and the $  \Delta^{\mathrm{F^2_P}} $ contribution in Table \ref{eH_HFS}. The evaluation of the resulting TPE correction is performed for the sum of elastic and inelastic contributions using the traditional expressions and the expressions based on the DRs for $ lp $ amplitudes. The latter leads to twice smaller uncertainties. However, the two evaluations agree within errors, which is a good test of the proton spin structure function $ g_2 $ parametrization for such calculation. The leading Zemach correction is a bit smaller than the result based on the typical proton form factors parametrization in Ref. \cite{Carlson:2008ke} due to the enhanced low-$ Q^2$ behavior of the magnetic proton form factor measured by the A1 Collaboration at MAMI \cite{Bernauer:2010wm,Bernauer:2013tpr}.

\begin{table}[H] 
\begin{center}
\begin{tabular}{|c|c|c|c|}
\hline
 & $ 10^6 \Delta $, $e \mathrm{H}$ &  $10^3 \Delta $, $\mu \mathrm{H}$    \\ \hline
Zemach, $ \Delta_{\mathrm{Z}} $ & $-$39.59(75) &  $-$7.36(14)   \\ 
Non-pole term, $\Delta^{\mathrm{F^2_P}} $ & 22.53(7) & 1.11(1)  \\ 
Recoil, $ \Delta^{\mathrm{p}}_{\mathrm{R}} $ & 5.31(7) & 0.8476(84)  \\ \hline
Traditional elastic, $ \Delta^{\mathrm{el}}_{0}  $ & $-$34.29(75) &  $-$6.51(14)   \\ 
Total elastic, $ {\Delta}^{\mathrm{el}}_{\mathrm{HFS}} $ & $-$43.40(74) &  $-$7.03(14)  \\ \hline
Total $ \Delta_{\mathrm{HFS}} $, within VVCS DRs & $-$32.60(2.56) &  $-$6.18(49) \\ 
Total $ \Delta_{\mathrm{HFS}} $ within $ l p $ DRs & $-$32.80(1.56) & $-$6.22(29)  \\ \hline
\end{tabular}
\caption{Finite-size TPE correction to the hyperfine splitting of S energy levels in $e$H and $ \mu $H. The Fermi energy HFS is $ 5.86785 ~\mathrm{\mu eV}$ for the 1S level in $e$H and $ 182.4432 ~\mathrm{meV}$ for the 1S level in $ \mu$H.} \label{eH_HFS}
\end{center}
\end{table}

\section{Conclusions and outlook}
\label{conclusions}
In this work we have derived dispersion relations for the forward lepton-proton scattering TPE amplitudes. We have fully expressed the spin-dependent TPE amplitudes in terms of the inclusive total cross sections at the leading $ \alpha $ order or equivalently in terms of the proton spin SFs. With these relations we have found a new way to determine the $O(\alpha^5)$ TPE proton structure correction to the S-level HFS in the hydrogen-like atoms explicitly accounting for the helicity double spin-flip amplitude. The result for individual channel HFS contribution based on the unitarity and analyticity of the lepton-proton scattering amplitudes is distinct to the standard approach, e.g., the elastic TPE contribution to HFS differs by the finite-sizable correction. Only accounting for contributions from all possible channels through the Burkard-Cottingham sum rule the two methods agree. We have reevaluated the TPE correction to HFS in electronic and muonic hydrogen connecting the region with small photons virtualities, which is expressed through the first and third moments of the proton spin SFs, to the region with large photons virtualities, where the data (parametrizations) on the proton form factors and SFs exists. The resulting TPE correction is similar to the literature result with slightly larger uncertainties in the standard approach to HFS and slightly smaller uncertainties within our approach. In a subsequent paper, we plan to decrease the errors of the TPE correction to HFS in the muonic hydrogen.

\section{Acknowledgments}
We thank Vladimir Pascalutsa for useful discussions about general formalism and advices given during this work and manuscript preparation, Marc Vanderhaeghen for useful discussions and advises concerning the numerical evaluations, Carl Carlson for the discussions concerning literature. We thank Keith Griffioen, Sebastian Kuhn, Nevzat Guler and Jacob Ethier for providing us with results on the proton spin SFs. This work was supported by the Deutsche Forschungsgemeinschaft (DFG) through Collaborative Research Center ``The Low-Energy Frontier of the Standard Model'' (SFB 1044), and Graduate School ``Symmetry Breaking in Fundamental Interactions'' (DFG/GRK 1581).
\appendix

\section{Crossing in lepton-proton scattering}
\label{even_odd}

In order to establish the even-odd properties for the invariant amplitudes under the crossing $ \omega \to - \omega $, we first perform the crossing on the lepton line and relate the amplitudes of the lepton-proton scattering $ f^{l^- p} (\omega) $ in the physical region ($ \omega > 0$) to the amplitudes of the antilepton-proton scattering $ f^{l^+p} (-\omega) $ in the unphysical region ($\omega < 0$).  We write the general form of the amplitude as
\ber \label{general_form}
T_{h^\prime \lambda^\prime, h \lambda} \left(\omega \right) & = & \sum_{i=1}^3 A_i \left(\omega \right) \bar{u}(k,h^\prime) O_i u (k,h) \bar{N} (p, \lambda ') O_i N (p, \lambda), 
\eer
with $O = \left(1,\, \gamma^{\mu\nu}, \, \gamma^\mu\gamma_5 \right)$. We observe that after the replacement in the lepton line $ k \to -k$, the amplitude transforms to
\ber
T_{h^\prime \lambda^\prime, h \lambda}^c \left(\omega\right) & = & \sum_{i=1}^3 A_i\left(-\omega\right) \bar{u}(-k,-h^\prime) O_i u (-k,-h) \bar{N} (p, \lambda ') O_i N (p, \lambda).
\eer
We can rewrite the lepton spinor $ u $ in terms of the antilepton spinor $ v $ as $ u\left(-k,-h\right) =  \gamma^2 v^{*} \left(k,h \right) $, where we exploit the same form $ u $ for the antilepton spinor as only particles or antiparticles participate in interaction. The expression for the helicity amplitude is given by
\ber
T_{h^\prime \lambda^\prime, h \lambda}^c \left(\omega \right) & = & \sum_{i=1}^3  A_i \left(-\omega \right) v^T \left(k, h^\prime \right) \gamma_2^+ \gamma_0 O_i  \gamma_2 v^{*} \left(k, h \right) \bar{N} (p, \lambda^\prime) O_i N (p, \lambda).
\eer
Transposing the lepton line we obtain:
\ber \label{crossed_form}
T_{h^\prime \lambda^\prime, h \lambda}^c \left(\omega \right) & = & \sum_{i=1}^3  A_i \left(-\omega \right) \bar{v} \left( k, h \right) \gamma_0 \left( \gamma_2^+ \gamma_0 O_i  \gamma_2 \right)^T v \left(k, h^\prime \right) \bar{N} (p, \lambda ') O_i N (p, \lambda). 
\eer
The tensor structure of Eq. (\ref{covariant_expression}) transforms to
\ber \label{covariant_expression_v}
T_{h^\prime \lambda^\prime, h \lambda}^c \left(\omega \right) & = & - \frac{f_+\left(-\omega \right)}{4 M m} \, \bar{v} \left( k,h \right) v \left(k,h^\prime \right)\,  \bar{N} (p, \lambda ')  N (p, \lambda) \nonumber\\
&-& \frac{m f_-\left(-\omega \right) - \omega g\left(-\omega \right)}{8 M \vec{k}^2} \, \bar{v} \left( k,h \right) \gamma^{\mu \nu} v \left(k,h^\prime \right)\,  \bar{N} (p, \lambda ')  \gamma_{\mu \nu} N (p, \lambda) \nonumber \\
&-& \frac{-\omega f_-\left(-\omega \right) + m g\left(-\omega \right) }{4 M \vec{k}^2 }  \, \bar{v} \left( k,h \right) \gamma_{\mu} \gamma_5 v \left(k,h^\prime \right) \, \bar{N} (p, \lambda ')  \gamma^{\mu} \gamma_5  N (p, \lambda),
\eer
which corresponds to the scattering of the antilepton off the proton.

 According to the crossing properties we can write amplitudes for the scattering of the antilepton $ f^{ l^+ p} $ in terms of the lepton scattering amplitudes  $ f^{ l^- p} $ as
\ber
f^{ l^+ p}_+\left(\omega  \right) & = & f^{ l^- p}_+\left( - \omega  \right), \\
f^{ l^+ p}_-\left(\omega  \right) & = & - f^{ l^- p}_-\left( - \omega  \right), \\
g^{ l^+ p}\left(\omega  \right) & = & g^{ l^- p}\left( - \omega  \right).
\eer

\section{Dispersion relations verification in QED}
\label{verification}

\begin{figure}[h]
\begin{center}
\includegraphics[width=.4\textwidth]{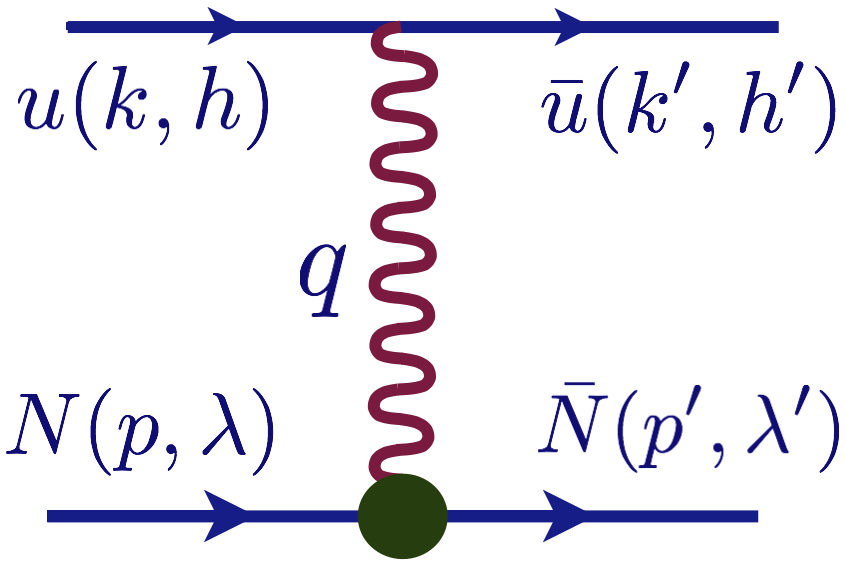}
\end{center}
\caption{One-photon exchange graph.}
\label{OPE_graph}
\end{figure}

In this Appendix, we verify the lepton-proton forward dispersion relations in QED. We reconstruct the real parts of the TPE amplitudes with the relations of Eqs. (\ref{M2_DR}-\ref{fermion_fermion_subtracted}) and compare them with the sum of the direct and crossed box graphs. The OPE helicity amplitude $ T^{1 \gamma}_{h^\prime \lambda^\prime h \lambda} $ for the lepton scattering off the charged point proton: $ l( k , h ) + p( p, \lambda ) \to l( k^\prime, h^\prime) + p(p^\prime, \lambda^\prime) $,  where $ h(h^\prime) $ denote the incoming (outgoing) lepton helicities and $ \lambda(\lambda^\prime) $ the corresponding proton helicities respectively, see Fig. \ref{OPE_graph}, is given by
\ber \label{1ph_QED}
T^{1 \gamma}_{h^\prime \lambda^\prime h \lambda} = \frac{e^2}{Q^2 + \mu^2} \bar{u} (k^\prime, h^\prime) \gamma^\mu u (k, h) \bar{N} (p^\prime , \lambda^\prime) \gamma_\mu N(p, \lambda) .
\eer
We introduce the finite photon mass $ \mu $ with the aim to have no deal with IR divergences. Such process is completely described by 2 Mandelstam variables, e.g., $ Q^2 = - (k-k^\prime)^2 $ - the squared momentum transfer, and $ s = ( p + k )^2 = M^2 + 2 M \omega + m^2 $ - the squared energy in the lepton-proton c.m. reference frame.

The relevant OPE cross sections are given by
\ber
\sigma^{1 \gamma}(\omega) & = & \frac{4 \pi M^2 \alpha^2}{\Sigma_s} \Bigg\{ \frac{\Sigma_s}{2M^2 s}  +  \left(\frac{s}{M^2}+\frac{4 \omega^2}{\mu ^2}+\frac{\mu ^2}{2 M^2}\right) \frac{\Sigma_s}{\Sigma_s+ s \mu ^2} \nonumber \\
&& -\frac{s + \mu^2}{M^2} \ln \frac{\Sigma_s + s \mu^2}{s \mu^2} \Bigg\}, \\
\sigma^{1 \gamma}_{\hp \hp}(\omega) - \sigma^{1 \gamma}_{\hp \hm}(\omega) & = & \frac{16 \pi M \alpha^2}{ \Sigma_s^2} \Bigg\{ \left( \omega \Sigma_s + 2 \mu^2 (m^2 + M \omega) ( \omega + M )  \right) \ln \frac{\Sigma_s + s \mu^2}{s \mu^2} \nonumber \\
&&  - \frac{ \Sigma_s \left(  \Sigma_s ( 2 \omega s - M \left( \omega^2 - m^2 \right)) + 2 s (m^2 + M \omega) ( \omega + M ) \mu^2 \right) }{s  (\Sigma_s + s \mu^2)} \Bigg\},   \\
\sigma^{1 \gamma}_{\perp}(\omega) - \sigma^{1 \gamma}_{\parallel}(\omega) & = & \frac{ 4 \pi \alpha^2 M m }{\Sigma_s^2} \Bigg\{  \frac{ 2 s \mu^2 \Sigma_s }{\Sigma_s + s \mu^2} +  \left( \Sigma_s - 2 s  \mu^2 \right) \ln \frac{\Sigma_s + s \mu^2}{s \mu^2} \Bigg\},
\eer
with $ \Sigma_s = \left( s - ( M + m )^2 \right) \left( s - ( M - m )^2 \right) = 4 M^2 \vec{k}^2$. The high-energy behavior of the relevant cross sections in the OPE approximation is as follows: 
\ber
\sigma^{1 \gamma} (\omega) \underset{\omega\gg}{  \sim} \omega^0 , \qquad \sigma^{1 \gamma}_{\hp \hp}(\omega) - \sigma^{1 \gamma}_{\hp \hm}(\omega) \underset{\omega\gg}{  \sim} \omega^{-1} \ln \omega,  \qquad \sigma^{1 \gamma}_{\perp} (\omega) - \sigma^{1 \gamma}_{\parallel} (\omega) \underset{\omega\gg}{  \sim} \omega^{-2} \ln \omega.
\eer 
The unsubtracted DR for the amplitude $ f_+^{2 \gamma} $ of Eq. (\ref{M1_DR}) is divergent as $ \omega $, therefore we use the subtracted DR of Eq. (\ref{fermion_fermion_subtracted}) for this amplitude.

The helicity amplitude corresponding with the TPE direct box graph $ T^{\mathrm{2 \gamma}}_{h^\prime \lambda^\prime h \lambda}$ is given by
\ber
\label{helamp_forward_TPE}
 T^{\mathrm{2 \gamma}}_{h^\prime \lambda^\prime h \lambda} =  \hspace{-0.1cm} \mathop{\mathlarger{\int}} \hspace{-0.15cm} \frac{  e^4 \mathrm{d}^4 q}{( 2 \pi )^4 i} \frac{\bar{u}(k,h^\prime) \gamma^\mu (\gamma.(k - q) + m) \gamma^\nu u (k,h) \bar{N}(p,\lambda^\prime) \gamma_\mu (\gamma.(p + q) + M) \gamma_\nu N (p,\lambda)}{\left(( p + q )^2 - M^2\right) \left(( k - q )^2 - m^2\right) \left(q^2 - \mu^2 \right) \left(q^2 - \mu^2 \right) }, \nonumber \\
\eer
with $\gamma. a  \equiv \gamma^\mu a_\mu$. We find the contribution of the direct box graph to the forward amplitudes $ f_+^{2 \gamma, \mathrm{dir}}, ~f_-^{2 \gamma, \mathrm{dir}} $ with exchange of two photons (see Fig. \ref{TPE_graph}) multiplying the fermion spinors by the spin projection operators. Then we sum over all possible polarizations evaluating traces of the Dirac matrices. The direct amplitudes $ f_+^{2 \gamma, \mathrm{dir}}, ~f_-^{2 \gamma, \mathrm{dir}} $ are given by
\ber
\label{helamp_forward_fp}
f_+^{2 \gamma, \mathrm{dir}} & = &  - 8 e^4 \frac{\partial}{\partial \mu^2}\mathop{\mathlarger{\int}}  \frac{  i \mathrm{d}^4 q}{( 2 \pi )^4} \frac{1}{( p + q )^2 - M^2} \frac{1}{( k - q )^2 - m^2}\frac{1}{q^2 - \mu^2 } \nonumber \\
& &  \times \left( 2 M^2 \nu_\gamma^2 - q^2 M \nu_\gamma  - \left( p \cdot q \right) \left( 2 M \nu_\gamma + m^2 +   \left( k \cdot q \right) \right) +\left( k \cdot q \right)( 2 M \nu_\gamma + M^2)  \right), \\
f_-^{2 \gamma, \mathrm{dir}} & = &  8 e^4 \frac{\partial}{\partial \mu^2}\mathop{\mathlarger{\int}}  \frac{  i \mathrm{d}^4 q}{( 2 \pi )^4} \frac{M}{( p + q )^2 - M^2} \frac{m}{( k - q )^2 - m^2}\frac{1}{q^2 - \mu^2 }   \left( q^2 \left( s \cdot S \right) - \left( q \cdot s \right) \left( q \cdot S \right)  \right), \nonumber \\
\eer
with the lepton and proton spin vectors in the laboratory frame $ s = (| \vec{k}|,~0,~0,~\omega)/m$ and $ S = (0,~0,~0,-1) $. For the double spin-flip amplitude $ g^{2 \gamma, \mathrm{dir}} $ we use the decomposition in terms of the scalar integrals, since the evaluation of traces cannot be exploited here due to the different spin directions of the initial and final fermions. Furthermore, we repeat the same steps for the crossed box graph contribution.

The optical theorem of Eqs. (\ref{optical_M_1}) and (\ref{optical_M_3}), the once-subtracted DR for $ f_+^{2 \gamma} $ amplitude of Eq. (\ref{fermion_fermion_subtracted}), the unsubtracted DRs for $  f_-^{2 \gamma},~ g^{2 \gamma} $ amplitudes of Eqs. (\ref{M2_DR}, \ref{M3_DR}) and the amplitudes properties under the crossing $ \omega \to - \omega $  of Eqs. (\ref{crossing_relations_TPE_1}-\ref{crossing_relations_TPE_3}) were checked comparing with the sum of the direct and crossed box graphs.

\section{Forward invariant amplitudes in terms of the proton structure functions}
\label{relations_between_amplitudes}

Substituting expressions for the inclusive cross sections of Eqs. (\ref{fp_cross_section}-\ref{g_cross_section}) into DRs, see Eqs. (\ref{M1_DR}-\ref{M3_DR}), we change the integration order and express the forward TPE amplitudes in terms of the proton SFs:
\ber 
\Re f^{2 \gamma}_+ \left(\omega\right) =  4 \alpha^2 M \int \limits^{\infty}_{0} \frac{\mathrm{d} Q^2}{Q^2} \int \limits^{\infty}_{\nu_{\mathrm{thr}}} \frac{\mathrm{d} \nu_\gamma}{\nu_\gamma} && \left\{ \frac{4 m^2 \nu_\gamma}{M Q^2} \left( 1 - 2 \tau_l \right) I^0_1  F_1\left( \nu_\gamma, Q^2 \right) \right. \nonumber \\
&& \left. + \left( I^0_1 + \frac{4 m \nu_\gamma}{Q^2} I^0_2 - \frac{I^0_3}{\tau_l} \right) F_2 \left( \nu_\gamma, Q^2 \right)  \right\},   \label{fplus_through_I} \\
 \Re f^{2 \gamma}_- \left(\omega\right) = 16 \alpha^2 \omega \int \limits^{\infty}_{0} \frac{\mathrm{d} Q^2}{Q^2} \int \limits^{\infty}_{\nu_{\mathrm{thr}}} \frac{\mathrm{d} \nu_\gamma}{\nu_\gamma} && \left
\{ \left( - I^0_1 +  \left( \tau_l + \tilde{\tau} \right) I^1_1 + \frac{\nu_\gamma}{m} \left( \frac{I^0_0}{2} + I^1_0  \right) \right) g_1\left( \nu_\gamma, Q^2 \right) \right. \nonumber \\
&& \left. + \frac{Q^2}{2 m \nu_\gamma} I^0_0 g_2 \left( \nu_\gamma, Q^2 \right)  \right\},  \label{fminus__through_I} \\
\Re g^{2 \gamma} \left(\omega\right) = 8 \alpha^2 m \int \limits^{\infty}_{0} \frac{\mathrm{d} Q^2}{Q^2} \int \limits^{\infty}_{\nu_{\mathrm{thr}}} \frac{\mathrm{d} \nu_\gamma}{\nu_\gamma} && \left
\{ \left(  I^0_1 +  \left( \tau_l + \tilde{\tau} \right) I^1_1 + \frac{\nu_\gamma}{m} \left( I^0_0 + I^1_0  \right) \right) g_1\left( \nu_\gamma, Q^2 \right) \right. \nonumber \\
&& \left. + 2 I^0_1 g_2 \left( \nu_\gamma, Q^2 \right)  \right\},  \label{g_through_I}
\eer
with the DR master integrals (we introduce the cut-off $ \Lambda $ in the divergent integrals):
\ber
I_0^0 & = &  \int \limits_{\omega_0}^{\infty} \frac{-m}{\omega^{\prime 2}-\omega^2} \frac{\mathrm{d} \omega^\prime}{\sqrt{\omega^{\prime 2} - m^2}} = \frac{m}{2 |\vec{k}| \omega} \ln \frac{\omega+|\vec{k}|}{\omega-|\vec{k}|} \frac{\omega |\vec{k}_0| - |\vec{k}| \omega_0}{\omega |\vec{k}_0| + |\vec{k}| \omega_0} , \label{I00} \\
I_1^0 & = &  \int \limits_{\omega_0}^{\infty} \frac{-\omega^\prime}{\omega^{\prime 2}-\omega^2} \frac{\mathrm{d} \omega^\prime}{\sqrt{\omega^{\prime 2} - m^2}} = \frac{1}{2 |\vec{k}|} \ln \frac{|\vec{k}|-|\vec{k}_0|}{|\vec{k}|+|\vec{k}_0|}, \label{I01} \\
I_2^0 & = &  \frac{1}{m} \int \limits_{\omega_0}^{\Lambda} \frac{-\omega^{\prime 2}}{\omega^{\prime 2}-\omega^2} \frac{\mathrm{d} \omega^\prime}{\sqrt{\omega^{\prime 2} - m^2}} = \frac{1}{m} \ln \frac{ 2 \Lambda}{\omega_0 + | \vec{k}_0 |} + \frac{\omega}{2 m |\vec{k}|} \ln \frac{\omega+|\vec{k}|}{\omega-|\vec{k}|} \frac{\omega |\vec{k}_0| - |\vec{k}| \omega_0}{\omega |\vec{k}_0| + |\vec{k}| \omega_0}, \label{I02}  \\
I_3^0 & = & \frac{1}{m^2} \int \limits_{\omega_0}^{\Lambda} \frac{-\omega^{\prime 3}}{\omega^{\prime 2}-\omega^2} \frac{\mathrm{d} \omega^\prime}{\sqrt{\omega^{\prime 2} - m^2}} = \frac{|\vec{k}_0| - \Lambda}{m^2} + \frac{\omega^2}{2 |\vec{k}| m^2} \ln \frac{|\vec{k}|-|\vec{k}_0|}{|\vec{k}|+|\vec{k}_0|}, \label{I03} \\
I_0^1 & = &  \int \limits_{\omega_0}^{\infty} \frac{-m^3}{\omega^{\prime 2}-\omega^2} \frac{\mathrm{d} \omega^\prime}{\left(\omega^{\prime 2} - m^2 \right)^{3/2}} = \frac{\omega_0 m}{|\vec{k}_0| \vec{k}^2} - \frac{m}{\vec{k}^2} + \frac{m^3}{2 |\vec{k}|^3 \omega} \ln \frac{\omega+|\vec{k}|}{\omega-|\vec{k}|} \frac{\omega |\vec{k}_0| - |\vec{k}| \omega_0}{\omega |\vec{k}_0| + |\vec{k}| \omega_0} , \label{I10} 
\eer
\ber
I_1^1 & = &  \int \limits_{\omega_0}^{\infty} \frac{-m^2 \omega^\prime}{\omega^{\prime 2}-\omega^2} \frac{\mathrm{d} \omega^\prime}{\left(\omega^{\prime 2} - m^2 \right)^{3/2}} = \frac{m^2}{\vec{k}^2 |\vec{k}_0|} + \frac{m^2}{2 |\vec{k}|^3} \ln \frac{|\vec{k}|-|\vec{k}_0|}{|\vec{k}|+|\vec{k}_0|}. \label{I11}
\eer

The reasonable result for the nucleon or narrow $ \Delta $ contribution, when we are allowed to interchange the $ \omega $ and $ Q^2, ~\nu_\gamma $ integration order, is given by the once-subtracted dispersion relation of Eq. (\ref{fermion_fermion_subtracted}). It can be formally obtained from the subtracted at the point $ \omega_s$ Eq. (\ref{fplus_through_I}) by $ \Re f^{2 \gamma}_+ \left(\omega\right) - \Re f^{2 \gamma}_+ \left(\omega_s\right)$:
\ber 
\Re f^{2 \gamma}_+ \left(\omega\right) & - & \Re f^{2 \gamma}_+ \left(m\right)  = \frac{4 m \alpha^2}{|\vec{k}|}  \nonumber \\
&& \times \int \limits^{\infty}_{0} \frac{\mathrm{d} Q^2}{Q^2} \int \limits^{\infty}_{\nu_{\mathrm{thr}}} \frac{\mathrm{d} \nu_\gamma}{\nu_\gamma} \left(  \frac{  \nu_\gamma \left( 1 - 2 \tau_l \right) }{2 m \tau_l}   \left( \ln \frac{|\vec{k}|-|\vec{k}_0|}{|\vec{k}|+|\vec{k}_0|} + \frac{2 | \vec{k}|}{|\vec{k}_0|}\right) F_1\left( \nu_\gamma, Q^2 \right) \right. \nonumber \\
&& \left.  + \frac{2 M \nu_\gamma }{m Q^2}   \left( \omega \ln \frac{ \left( \omega+|\vec{k}| \right)^2 \left(\omega_0^2 - \omega^2 \right)}{ \left( \omega |\vec{k}_0| + |\vec{k}| \omega_0 \right)^2 } + 2 \left(\omega_0 - |\vec{k}_0| \right) \frac{|\vec{k}|}{|\vec{k}_0|} \right) F_2 \left( \nu_\gamma, Q^2 \right)   \right. \nonumber \\
&& \left. - \frac{M}{m} \left( \frac{  \omega^2 - m^2 \tau_l }{2 m^2 \tau_l}  \ln \frac{|\vec{k}|-|\vec{k}_0|}{|\vec{k}|+|\vec{k}_0|} + \frac{1 -  \tau_l}{\tau_l}  \frac{|\vec{k}|}{|\vec{k}_0|} \right)  F_2 \left( \nu_\gamma, Q^2 \right) \right).
\eer
The $ \nu_\gamma $ integrals for the spin-dependent amplitudes of Eqs. (\ref{M3}) and (\ref{M2}) are convergent.

Due to the Regge behavior of the $ F_1 $ proton structure function given by the Pomeron exchange, the $ \nu_\gamma $ integrals are divergent, and the DR for the amplitude $ f^{2 \gamma}_+$ is not applicable for the inelastic intermediate states TPE contribution.

\section{HFS through the forward double virtual Compton scattering amplitudes}
\label{box_graph_method}

\begin{figure}[h]
\begin{center}
\includegraphics[width=.6\textwidth]{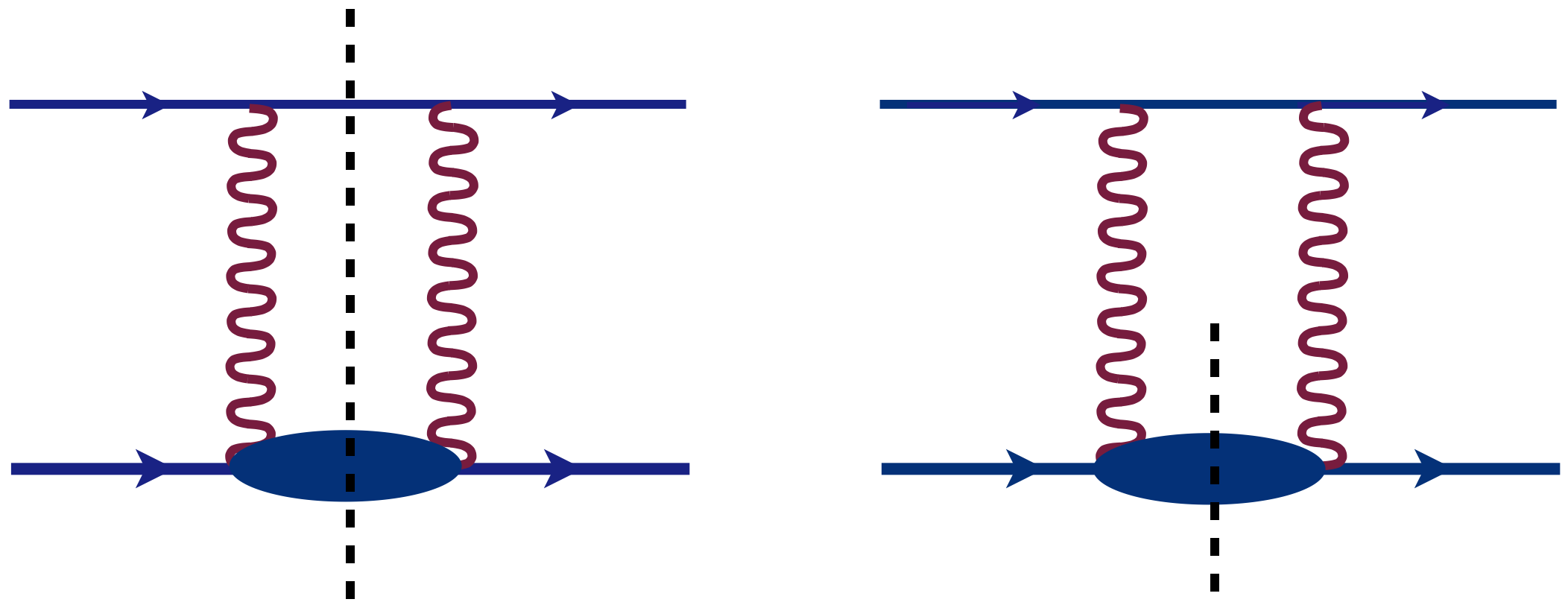}
\end{center}
\caption{Forward elastic $ l p $ scattering with cut of both fermion lines (left panel) and with cut of the nucleon line only (right panel).}
\label{box_diagram_cuts}
\end{figure}

It is instructive to evaluate the HFS correction cutting only the lower blob of the TPE graph \cite{Carlson:2008ke,Carlson:2011af}, see right panel in Fig. \ref{box_diagram_cuts}.

\begin{figure}[h]
\begin{center}
\includegraphics[width=.45\textwidth]{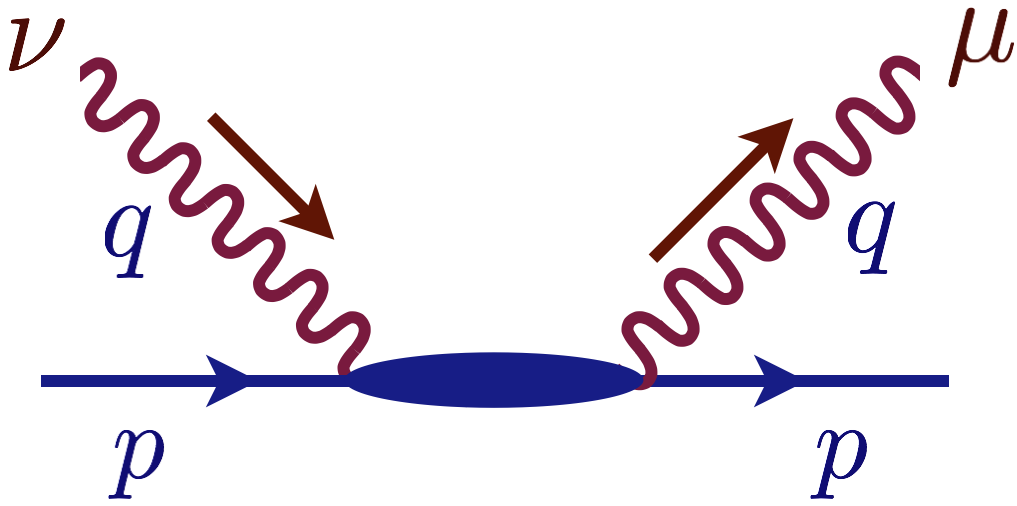}
\end{center}
\caption{Forward VVCS process.}
\label{VVCS_forward}
\end{figure}

The forward TPE amplitude can be evaluated considering the lower blob of the TPE graph in Fig. \ref{TPE_graph} as a forward VVCS process on a proton: $ \gamma^\ast \left(q, \lambda_1\right) + N\left(p,\lambda\right) \to \gamma^\ast\left(q, \lambda_2\right) + N\left(p, \lambda^\prime\right) $, which is shown in Fig. \ref{VVCS_forward}. 
The forward VVCS amplitude $ T_{\lambda_2 \lambda^\prime, \lambda_1 \lambda} $ can be written in terms of the forward VVCS tensor $ M^{\mu \nu} $ as
\ber \label{forward_vvcs_helamp}
T_{\lambda_2 \lambda', \lambda_1 \lambda} = \varepsilon_\nu\left(q, \lambda_1\right) \varepsilon^\ast_\mu\left(q, \lambda_2\right) \cdot  \bar{N}\left(p, \lambda'\right) (4 \pi  M^{\mu \nu})  N\left(p, \lambda\right),
\eer
where $ \varepsilon_\nu, ~\varepsilon^\ast_\mu $ denote the virtual photon polarization vectors, $ N, \bar{N} $ the proton spinors, and  $ \lambda_1, \lambda_2 ~(\lambda, \lambda^\prime) $ the photon (proton) helicities.

The forward VVCS tensor $ M^{\mu \nu} $ can be expressed as a sum of the symmetric (spin-independent) $ M^{\mu \nu}_S $ and the antisymmetric (spin-dependent) $ M^{\mu \nu}_A $ parts:
\ber \label{forward_vvcs_tensor}
M^{\mu \nu} & = & M^{\mu \nu}_S + M^{\mu \nu}_A, \\
M^{\mu \nu}_S & = & \left( - g^{\mu\nu} + \frac{q^{\mu}q^{\nu}}{q^2}\right)
\mathrm{T}_1 (\nu_\gamma, Q^2) + \frac{1}{M^2} \left(p^{\mu}-\frac{ \left(p\cdot
q \right)}{q^2}\,q^{\mu}\right) \left(p^{\nu}-\frac{\left( p\cdot
q \right)}{q^2}\, q^{\nu} \right) \mathrm{T}_2 (\nu_\gamma, Q^2), \nonumber \\  \label{forward_vvcs_tensor_sym} 
\eer
\ber
M^{\mu \nu}_A & = &  \frac{1}{2 M^2} \bigg [ M \{ \gamma^{\mu \nu}, \gamma.q \} \mathrm{S}_1 (\nu_\gamma, Q^2) +  \Big (  [ \gamma^{\mu}, \gamma^{\nu} ] q^2 + q^\mu [\gamma^\nu,  \gamma.q]  + q^\nu [ \gamma.q,  \gamma^\mu]  \Big) \mathrm{S}_2 (\nu_\gamma, Q^2) \bigg ],\nonumber \\  \label{forward_vvcs_tensor_asym}
\eer
with the forward Compton amplitudes $ \mathrm{T}_1,~\mathrm{T}_2, ~\mathrm{S}_1,~\mathrm{S}_2 $, which are functions of $ \nu_\gamma,~Q^2 $ and enter Eq. (\ref{forward_vvcs_tensor}) in a gauge-invariant way, e.g., $ q_\mu M^{\mu \nu} = q_\nu M^{\mu \nu} = 0$. The optical theorem relates the imaginary parts of the forward Compton amplitudes to the proton SFs by
\ber
 \Im \mathrm{T}_1\left(\nu_\gamma, Q^2\right) & = & \pi F_1\left(\nu_\gamma, Q^2\right), \quad~~~  \Im \mathrm{T}_2\left(\nu_\gamma, Q^2\right)  =  \frac{\pi M}{\nu_\gamma} F_2\left(\nu_\gamma, Q^2\right), \\
 \Im \mathrm{S}_1\left(\nu_\gamma, Q^2\right) & = & \frac{\pi M}{\nu_\gamma} g_1\left(\nu_\gamma, Q^2\right), \quad  \Im \mathrm{S}_2 \left(\nu_\gamma, Q^2\right)  =  \frac{\pi M^2}{\nu_\gamma^2} g_2\left(\nu_\gamma, Q^2\right).
\eer

The forward lepton-proton scattering TPE amplitude can be expressed in terms of the forward VVCS amplitude $ M^{\mu \nu}$ as
\ber \label{TPE_amplitude_through_VVCS}
 T^{2 \gamma}_{h^\prime \lambda^\prime h \lambda} \left( \omega \right) & = &   e^2  \mathop{\mathlarger{\int}} \frac{i \mathrm{d}^4 q}{\left( 2 \pi \right)^3} \frac{ \tilde{L}^{\mu \nu }_{h^\prime h}  \bar{N}(p,\lambda') M_{\mu \nu} N(p,\lambda)}{\left(q^2 \right)^2} ,
\eer
with the leptonic tensor $ \tilde{L}^{\mu \nu }_{h^\prime h} $:
\ber
\tilde{L}^{\mu \nu }_{h^\prime h} & = & \bar{u}\left(k,h'\right)  \left( \gamma^\mu \frac{\hat{k}-\hat{q}+m}{\left(k-q\right)^2-m^2} \gamma^\nu + \gamma^\nu \frac{\hat{k}+\hat{q}+m}{\left(k+q\right)^2-m^2} \gamma^\mu  \right) u\left(k,h\right).
\eer

The expression for the spin-independent forward TPE amplitude $ f_+^{2\gamma} $ is given by
\ber
f^{2\gamma}_+ \left(\omega \right) =  - 4 e^4 \mathop{\mathlarger{\int}}  \frac{i  \mathrm{d}^4 q}{( 2 \pi )^4} \frac{ \left( k \cdot q \right)^2 \left( 2 \mathrm{T}_1 - \mathrm{T}_2 \right) + q^2 \left( m^2 \mathrm{T}_1 - \omega^2 \mathrm{T}_2 \right) + \frac{2 \omega}{M} \left( k \cdot q \right) \left( p \cdot q \right) \mathrm{T}_2}{\left( q^4 - 4 \left(k \cdot q \right)^2 \right) (q^2)^2 }. 
\eer
This result at threshold is in agreement with Ref. \cite{Carlson:2011zd}.

The expressions for the spin-dependent forward TPE amplitudes $ f_-^{2\gamma}, ~g^{2\gamma} $ are given by
\ber
f^{2\gamma}_- \left(\omega \right)& = & - \frac{8 \alpha}{M^3} \mathop{\mathlarger{\int}}  \frac{i \mathrm{d}^4 q}{ \pi^2} \frac{ \left( M^2 q^2  +  \frac{M^2 \left(k \cdot q \right)^2 + m^2 \left(p \cdot q \right)^2 - 2 \left( k \cdot p \right) \left(k \cdot q \right) \left(p \cdot q \right) }{\omega^2 - m^2} \right) \left( k \cdot p \right)  \mathrm{S}_1 + M^2  q^2 \left( k \cdot q \right)  \mathrm{S}_2 }{\left( q^4 - 4 \left(k \cdot q \right)^2 \right) q^2} \nonumber \\
&&  - \frac{8 \alpha}{M} \mathop{\mathlarger{\int}}  \frac{i \mathrm{d}^4 q}{ \pi^2}  \frac{ \left( k \cdot q \right) \left( p \cdot q \right)  \mathrm{S}_1}{\left( q^4 - 4 \left(k \cdot q \right)^2 \right) q^2} , \\
g^{2\gamma} \left(\omega \right)& = & \frac{4 m \alpha }{M^2} \mathop{\mathlarger{\int}}  \frac{  i \mathrm{d}^4 q}{\pi^2} \frac{ \left( M^2 q^2 - \frac{M^2 \left(k \cdot q \right)^2 + m^2 \left(p \cdot q \right)^2 - 2\left( k \cdot p \right) \left(k \cdot q \right) \left(p \cdot q \right) }{\omega^2 - m^2} \right)  \mathrm{S}_1 + 2 q^2 \left( p \cdot q \right)  \mathrm{S}_2}{ \left( q^4 - 4 \left(k \cdot q \right)^2 \right) q^2 }.
\eer
Evaluating these expressions in the nucleon rest frame at threshold we obtain:
\ber
  g^{2 \gamma}(m) & = & - f^{2 \gamma}_-(m) = \frac{8 m }{M^2} e^4 \mathop{\mathlarger{\int}}  \frac{ i \mathrm{d}^4 q}{( 2 \pi )^4} \frac{ q^2 \nu_\gamma \mathrm{S}_2 + M \frac{2 q^2 + \nu_\gamma^2}{3} \mathrm{S}_1 }{ \left( q^4 - 4 \left(k \cdot q \right)^2 \right) q^2 },
\eer
in agreement with Eq. (\ref{Old_result_condition}) and the standard HFS derivation of Eq. (\ref{2S_correction_oldxx})  \cite{Iddings:1959zz,Iddings:1965zz,Drell:1966kk,Faustov:1966,Faustov:1970,Bodwin:1987mj}. 

Exploiting DRs for the spin-dependent Compton amplitudes $ \mathrm{S}_1 $ and $ \mathrm{S}_2 $:
\ber
\Re \mathrm{S}_1 \left(\nu_\gamma, Q^2\right)  & = &  \mathop{\mathlarger{\int}} \limits^{~~ \infty}_{\nu_{\mathrm{thr}}} \frac{ 2 M g_1 \left(\nu_\gamma', Q^2\right)}{\nu_\gamma'^2-\nu_\gamma^2 - i \varepsilon }  \mathrm{d} \nu_\gamma',\\
\Re \mathrm{S}_2 \left(\nu_\gamma, Q^2\right)  & = &  \mathop{\mathlarger{\int}} \limits^{~~ \infty}_{\nu_{\mathrm{thr}}} \frac{ 2 \nu_\gamma M^2 g_2 \left(\nu_\gamma', Q^2\right)}{\nu_\gamma'^2 \left( \nu_\gamma'^2-\nu_\gamma^2 - i \varepsilon \right) }  \mathrm{d} \nu_\gamma',
\eer
we obtain the same expression for the amplitude $ f^{2 \gamma}_- $, see Eq. (\ref{fminus}), as with the DRs for the forward $lp$ amplitudes. The difference in the amplitude $ g^{2 \gamma} $, see Eq. (\ref{g_ampl}), is given by the $ \omega $-independent term, which is just the constant real part in $g^{2 \gamma} $ amplitude. It vanishes with account of the BC sum rule. However, if one uses the DR for the amplitude $ \nu_\gamma \mathrm{S}_2 $ \cite{Hagelstein:2015egb}, the result for $ f^{2 \gamma}_- (m) $ coincides with the expression $ - g^{2 \gamma} (m) $.

The forward TPE amplitudes evaluated within the DR approach coincide with the amplitudes evaluated with a help of the DRs for the forward VVCS amplitudes. However, the contribution of an individual TPE intermediate state differs in these two approaches.

\section{Forward scattering observables}
\label{scattering_observables}

The forward unpolarized elastic scattering cross section in the c.m. reference frame is given by
\ber
\frac{\mathrm{d} \sigma}{\mathrm{d} \Omega} \left( \theta = 0 \right) = \frac{|f_+(\omega)|^2 + |f_-(\omega)|^2 + 2 |g(\omega)|^2}{64 \pi^2 ( M^2 + 2 M \omega + m^2)},
\eer
with the electron scattering angles $\Omega$ and the azimuthal angle $ \theta$.

All possible single-spin asymmetries are zero for the scattering in the forward direction. We denote the lepton spin asymmetry for the scattering on the polarized proton as $ A $ and for the scattering on the unpolarized proton with the polarization transfer to the final proton as $ P $. The asymmetries for the longitudinally polarized lepton and the longitudinally polarized proton in the forward scattering are expressed as
\ber
 A_l & = & \frac{\mathrm{d} \sigma_{+-} - \mathrm{d} \sigma_{--}}{\mathrm{d} \sigma_{+-} + \mathrm{d} \sigma_{--}} = - 2 \frac{ \Re \left(f_+ f_-^*\right) + |g|^2 }{|f_+|^2 + |f_-|^2 + 2 |g|^2}, \\
 P_l & = & \frac{\mathrm{d} \sigma_{+-} - \mathrm{d} \sigma_{--}}{\mathrm{d} \sigma_{+-} + \mathrm{d} \sigma_{--}} = - 2 \frac{  \Re \left(f_+ f_-^*\right) -  |g|^2 }{|f_+|^2 + |f_-|^2 + 2 |g|^2} .
\eer 
The asymmetries for the transversely polarized lepton and the transversely polarized proton in the forward scattering are expressed as
\ber
 A_t & = & \frac{\mathrm{d} \sigma_{\uparrow \uparrow} - \mathrm{d} \sigma_{\uparrow \downarrow}}{\mathrm{d} \sigma_{\uparrow \uparrow} + \mathrm{d} \sigma_{\uparrow \downarrow}} = 2 \frac{ \Re \left(\left(f_+ + f_-\right) g^* \right) }{|f_+|^2 + |f_-|^2 + 2 |g|^2},  \\
 P_t & = & \frac{\mathrm{d} \sigma_{\uparrow \uparrow} - \mathrm{d} \sigma_{\uparrow \downarrow}}{\mathrm{d} \sigma_{\uparrow \uparrow} + \mathrm{d} \sigma_{\uparrow \downarrow}} = 2 \frac{ \Re \left(\left(f_+ - f_-\right) g^* \right) }{|f_+|^2 + |f_-|^2 + 2 |g|^2} .
\eer
The asymmetries in the case of one transverse and one longitudinal polarizations vanish.

\end{document}